\documentclass[preprint,3p,sort&compress]{elsarticle}
\usepackage{amsmath,amssymb}
\usepackage{epsfig}
\usepackage{color}
\usepackage{verbatim}
\usepackage[normalem]{ulem}
\usepackage[utf8]{inputenc}

\def\gs{\mathrel{
   \rlap{\raise 0.511ex \hbox{$>$}}{\lower 0.511ex \hbox{$\sim$}}}}
\def\ls{\mathrel{
   \rlap{\raise 0.511ex \hbox{$<$}}{\lower 0.511ex \hbox{$\sim$}}}}

\newcommand{\ba}{\begin{array}{c}}
\newcommand{\baz}{\begin{array}{cc}}
\newcommand{\bad}{\begin{array}{ccc}}
\newcommand{\bav}{\begin{array}{cccc}}
\newcommand{\baf}{\begin{array}{ccccc}}
\newcommand{\bena}{\begin{eqnarray}}
\newcommand{\eena}{\end{eqnarray}}
\newcommand{\bea}{\begin{equation} \begin{array}{c}}
\newcommand{\eea}{ \end{array} \end{equation}}
\newcommand{\ea}{\end{array}}

\journal{Astroparticle Physics}

\usepackage{amsmath,amssymb}

\begin{document}

\begin{flushright}
  FTUAM-16-30 \\
  IFT-UAM/CSIC-16-078
\end{flushright}

\title{Prospects for Detecting Galactic Sources of Cosmic Neutrinos with IceCube: An Update}

\author[wipac]{Francis Halzen}
\ead{francis.halzen@icecube.wisc.edu}

\author[wipac]{Ali Kheirandish}
\ead{ali.kheirandish@icecube.wisc.edu}

\address[wipac]{
Wisconsin IceCube Particle Astrophysics Center and Department of Physics,
University of Wisconsin, Madison, WI 53706, USA
}

\author[uam_ift]{Viviana Niro}
\ead{viviana.niro@uam.es}

\address[uam_ift]{
Departamento de F\'isica Te\'orica, Universidad Aut\'onoma de Madrid, and Instituto 
de F\'isica Te\'orica UAM/CSIC, Calle Nicol\'as Cabrera 13-15, Cantoblanco, E-28049 Madrid, Spain
}

\begin{abstract}
Air-Cherenkov telescopes have mapped the Galactic plane at TeV energies. Here we evaluate the prospects for detecting the neutrino emission from sources in the Galactic plane assuming that the highest energy photons originate from the decay of pions, which yields a straightforward prediction for the neutrino flux from the decay of the associated production of charged pions. Four promising sources are identified based on having a large flux and a flat spectrum. We subsequently evaluate the probability of their identification above the atmospheric neutrino background in IceCube data as a function of time. We show that  observing them over the twenty-year lifetime of the instrumentation is likely, and that some should be observable at the $3\,\sigma$ level with six years of data. 
In the absence of positive results, we derive constraints on the spectral index and cut-off energy of the sources, 
assuming a hadronic acceleration mechanism.
\end{abstract}


\maketitle

\section{\label{sec:intro} Introduction}
The IceCube experiment has discovered a flux of high-energy neutrinos from extragalactic sources with an energy density similar to that observed for gamma rays ~\cite{halzen:2016}. The observation underscores the important role of sources  
accelerating protons that produce similar energy in photons and neutrinos, which are the decay products of neutral and charged pions, respectively. 

The present data cannot exclude a subdominant flux of Galactic origin in the IceCube data~\cite{Taylor:2014hya,Gaggero:2015xza,Ahlers:2015moa,Palladino:2016zoe}. 
Unidentified sources \cite{Fox:2013oza}, 
Fermi bubbles~\cite{Taylor:2014hya,Lunardini:2011br,Lunardini:2013gva}, and Sagittarius~A$^*$~\cite{Bai:2014kba} have 
been reviewed as potential Galactic sources. However, the general conclusion is that these sources 
can account for a fraction of the events detected. Specifically, the possibility that the hot spot close to the Galactic Center (GC) is produced by a single point 
source with a flux normalization of $6~\times~10^{-8}~\rm{GeV}~\rm{cm}^{-2}~\rm{s}^{-1}$ has been 
excluded~\cite{Adrian-Martinez:2014wzf,Gonzalez-Garcia:2013iha}. 

In a map of the northern Galactic plane obtained with Milagro data, six promising neutrino sources were identified in Refs.~\cite{Halzen:2008zj,GonzalezGarcia:2009jc}. 
The IceCube Collaboration has carried out extensive searches for point and extended sources in Ref.~\cite{Aartsen:2014cva}, 
reporting evidence with a significance of $2.5~\sigma$, when the six Milagro sources are considered together~\cite{Aartsen:2014cva}. 

In the previous study, Ref.~\cite{Gonzalez-Garcia:2013iha}, the authors revisited the prospects for observing the three confirmed Milagro sources and re-evaluated the probability and constraints in light of the low-energy cut-off reported by the Milagro collaboration \cite{Abdo:2012jg,Smith:2010yn}. The authors concluded that more than 10 years of running IceCube is necessary to yield a discovery at the level of $3\sigma$. In the case of the source MGRO J1908+06, evidence at $3\sigma$ could be obtained in seven years assuming values 
of the spectral index and the cut-off energy that are in good agreement with the best fit reported in~\cite{Abdo:2012jg}. 

In this paper, we will update the theoretical predictions using the observation and flux measurements reported by HAWC, ARGO-YBJ, and  air Cherenkov telescopes (ACT) VERITAS and HESS. Most importantly, with a detector superior to Milagro, the HAWC experiment has confirmed only four of the six sources~\cite{Abeysekara:2015qba,hawc_gamma}:  MGRO J1908+06, MGRO J1852+01, MGRO J2031+41, and MGRO J2019+37. For these, we will construct a gamma ray spectrum based on all information available and evaluate the neutrino flux. Subsequently, we will compute the number of signal and background events as well as the p-value for observing the sources as a function of time. Finally, we will determine exclusion limits on a flux of hadronic origin in the absence of an observation. Our main results can be summarized as follows: 

\begin{itemize}
\item MGRO J1908+06: Although historically classified as a pulsar wind nebula (PWN) and currently as an unidentified source, its large size and hard spectrum in TeV photons suggest that it may be a supernova remnant (SNR). SNRs are suspected to be the sources of the highest energy cosmic rays in the Galaxy. We re-evaluate the probability of observing the source using the flux reported by HESS and anticipate a $3\sigma$ observation in about 10 years of IceCube data. However, the answer depends on the actual threshold of the specific analysis. By increasing the energy threshold, IceCube has the potential to observe MGRO J1908+06 at the some statistical level with only six years of data. A lack of observation in 15 years of IceCube data will indicate that MGRO J1908+06 is not a cosmic-ray accelerator.

\item MGRO J1852+01: In the original Milagro map of the TeV sky, this source missed the statistical threshold for candidate sources. It has now been conclusively observed by HAWC and is a potential neutrino source considering its relatively large flux. Since the proper study of spectrum and extension of the source have not been performed by HAWC, we have studied the neutrino flux under different assumptions for the source's extension and spectrum. We find that IceCube should see this source in 5 years of data provided that the source is not extended. However, if the source is extended, 15 years is required to reach a significant level of observation.

\item MGRO J2031+41: Due to the uncertainties associated with the origin of the flux of the Cygnus cocoon and $\gamma$-Cygni, a complete picture of this source is missing. Its extension and other TeV emissions in its vicinity have made it difficult for ACT experiments like VERITAS to measure the TeV flux from this source. Although previous studies indicated that observing the source would be challenging~\cite{Gonzalez-Garcia:2013iha}, using recent ARGO-YBJ and Fermi data, we argue that neutrino observations at the level of $3\sigma$ may be possible in 10 years of IceCube data.

\item MGRO J2019+37: 
We present an update on the neutrino observation from this source based on the spectrum measured by VERITAS, which has provided up to now the most precise measurement for the spectrum of the source up to 30 TeV.
We show that IceCube is likely to observe the source in 15 years. This source is currently classified as a PWN. Thus, the detection of 
neutrinos from this region could point towards the production mechanism of neutrinos in a PWN as described in Ref. \cite{Lemoine:2014ala}.

\end{itemize}

\section{\label{sec:Milagro} Milagro sources}
After confirmation by HAWC ~\cite{Abeysekara:2015qba,hawc_gamma}, the Milagro sources that we consider in this analysis are, as mentioned above, MGRO J1908+06, MGRO J1852+01, MGRO J2031+41, and MGRO J2019+37. In this section, we summarize the experimental information on these sources. \\

\underline{\textbf{MGRO J1908+06:}} 
The source MGRO~J1908+06 has been detected by large-acceptance air-shower detectors (EAS) like the Milagro experiment, see Refs.~\cite{Abdo:2007ad,Abdo:2009ku,Smith:2010yn}, 
and the ARGO-YBJ experiment~\cite{ARGO-YBJ:2012goa}. 
This source has been detected also by ACTs, like HESS~\cite{Aharonian:2009je}, which 
finds a spectrum with no evidence of a cut-off for energies $<20$~TeV. 
The HESS detector reports a flux systematically lower than the Milagro and ARGO-YBJ data. With better angular resolution, it could be that HESS detects the flux from a point source that is not resolved by the Milagro and ARGO-YBJ observation. MGRO J1908+06 has also been recently detected by VERITAS~\cite{Aliu:2014xra}, and the flux reported is of the same order as the one measured by HESS. Also, the value recently reported by HAWC points towards 
a similar normalization~\cite{Abeysekara:2015qba}.  

\begin{figure}[!t]
\center
\begin{tabular}{rl}
\includegraphics[width=0.48\textwidth]{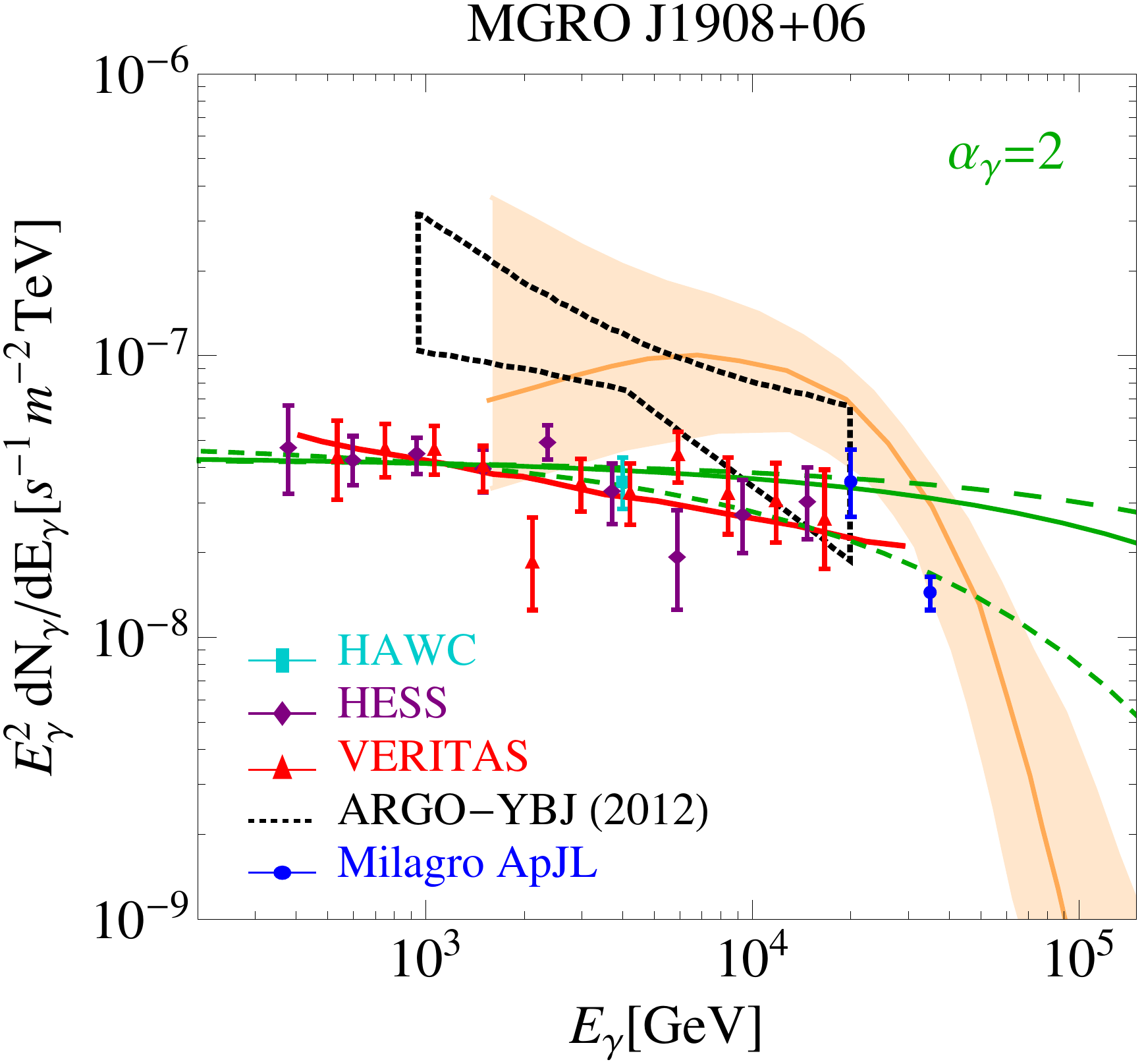} &
\includegraphics[width=0.48\textwidth]{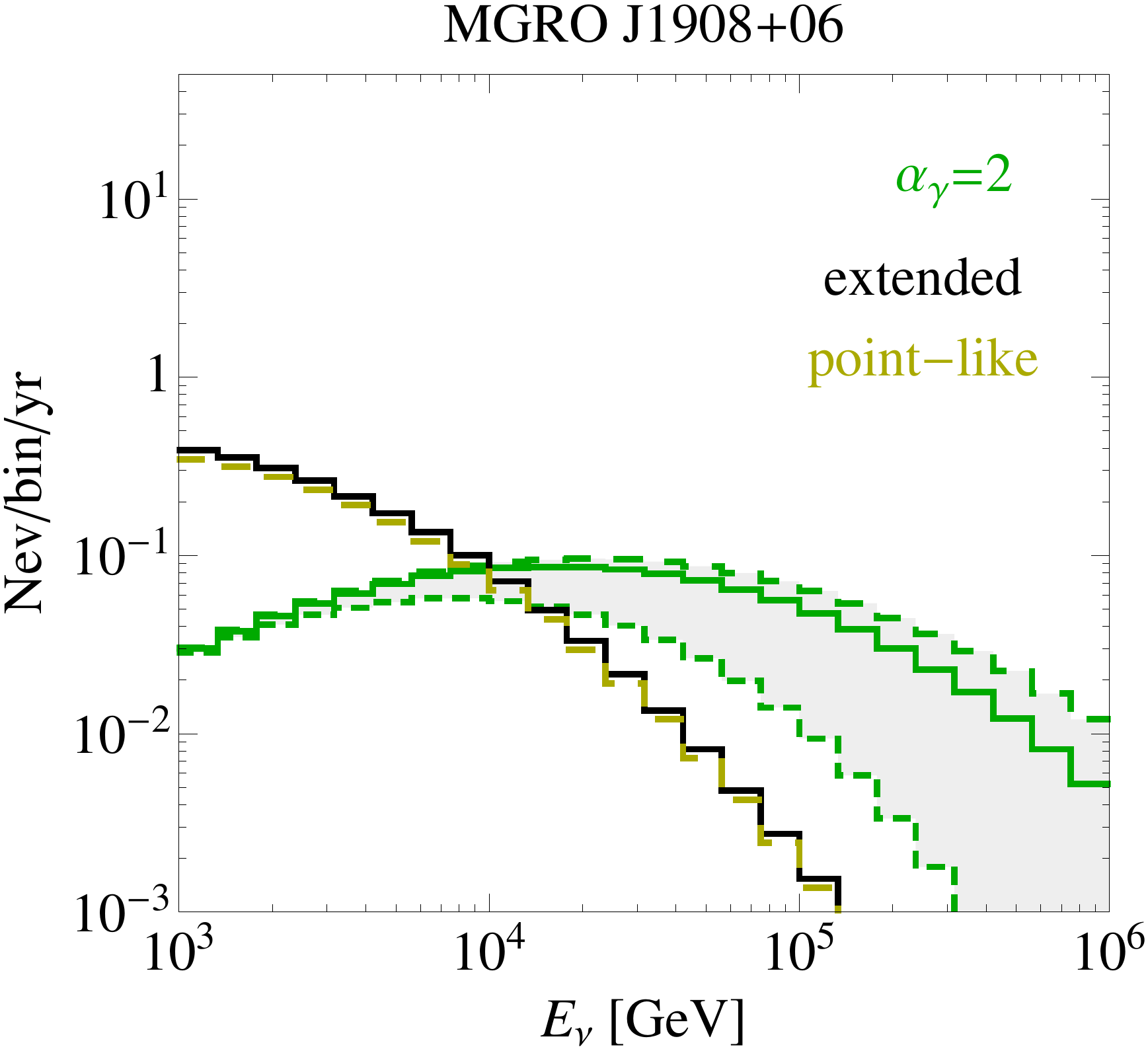} 
\end{tabular}
\caption{\label{fig:sources_spectra_second} 
{\it \underline{Left panel:}} 
We show in purple the data by HESS~\cite{Aharonian:2009je}, in red the one from 
VERITAS~\cite{Aliu:2014rha}, and in cyan the one from HAWC~\cite{Abeysekara:2015qba}.  
In blue we show the previous flux measurements by
Milagro~\cite{Abdo:2007ad,Abdo:2009ku}, while the solid
orange line and the shaded orange area show the best fit and the
$1\sigma$ band as reported in Ref.~\cite{Smith:2010yn} by Milagro. 
The dotted area is the
ARGO-YBJ $1\sigma$ band~\cite{ARGO-YBJ:2012goa}. 
With green lines we show the spectra obtained considering 
$\alpha_\gamma=2$ and fixing the normalization to the best fit reported in 
Table~\ref{tab:sources_fit}, where we also allowed the cut-off energy to vary: 
$E_{\rm cut, \gamma} =30,~300, and~800$~TeV (short-dashed, solid, and long-dashed lines, in green). 
{\it \underline{Right panel:}} We show the corresponding number of events for these spectra. 
The gray band encodes the uncertainty on the cut-off energy. 
With the black (gold dashed) line, we show the background from atmospheric neutrinos for extended (point-like) sources. 
}
\end{figure}

We report in Table~\ref{tab:sources_ext} the extension for MGRO J1908+06 observed by HESS, VERITAS, 
and ARGO-YBJ, while in Table~\ref{tab:sources_fit} we report the flux measured by HESS and VERITAS. 
In Fig.~\ref{fig:sources_spectra_second}, we have compiled the spectra for MGRO J1908+06 from the different experiments. 

Finally, note that Fermi-LAT observes the pulsar PSR J1907+0602 within the extension of the Milagro source MGRO J1908+06~\cite{Abdo:2010ht}. On the other hand, the large size and hard spectrum in TeV photons of MGRO J1908+06 are not characterstic of a PWN and perhaps consistent with a SNR. SNRs are suspected to be the sources of the highest energy cosmic rays in the Galaxy~\cite{BaadeAndZwicky}, see also 
Ref.~\cite{Ackermann:2013wqa,Gabici:2007qb}.  
\\

\begin{figure}[!t]
\begin{tabular}{rl}
\includegraphics[width=0.48\textwidth]{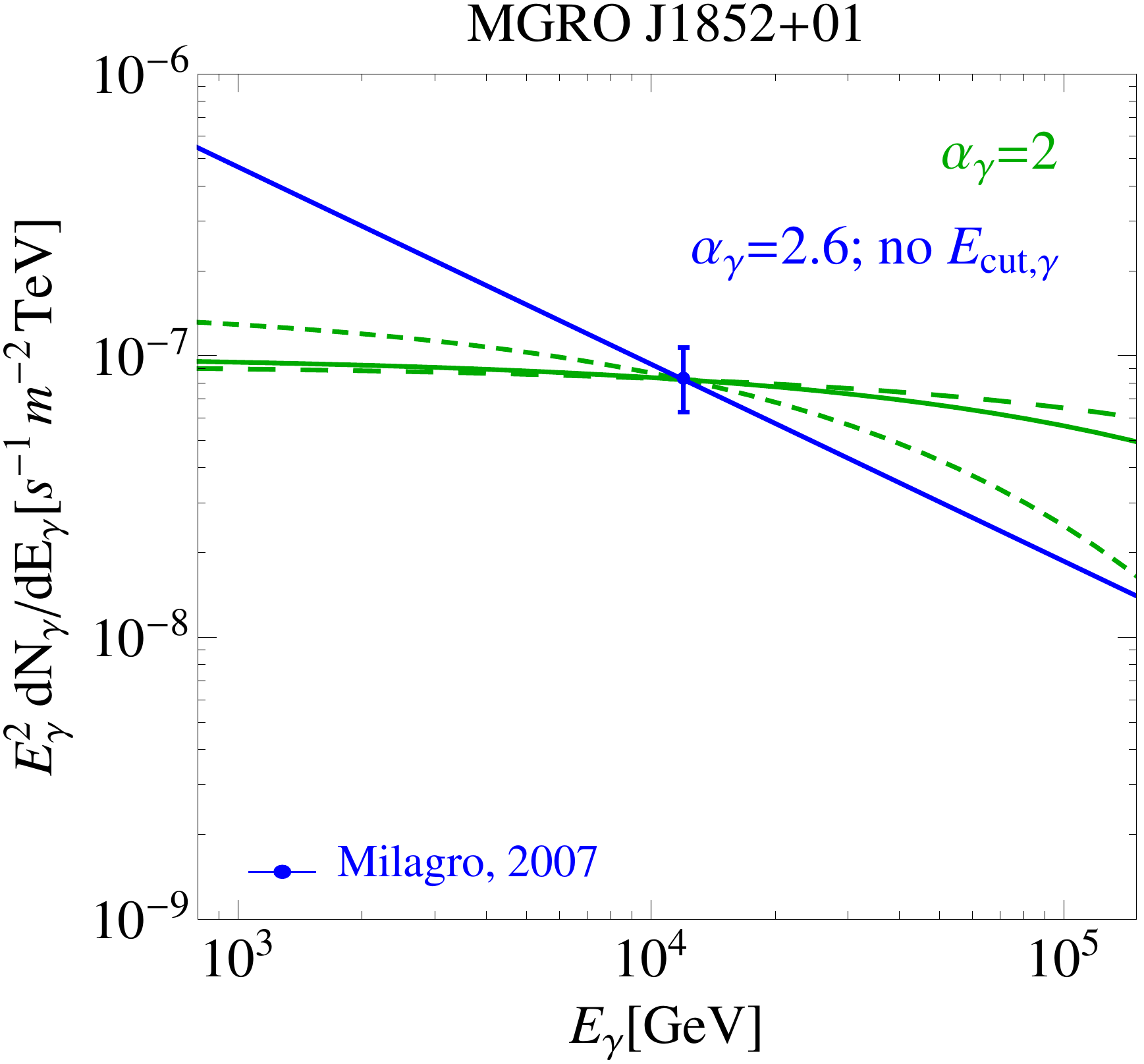} &
\includegraphics[width=0.48\textwidth]{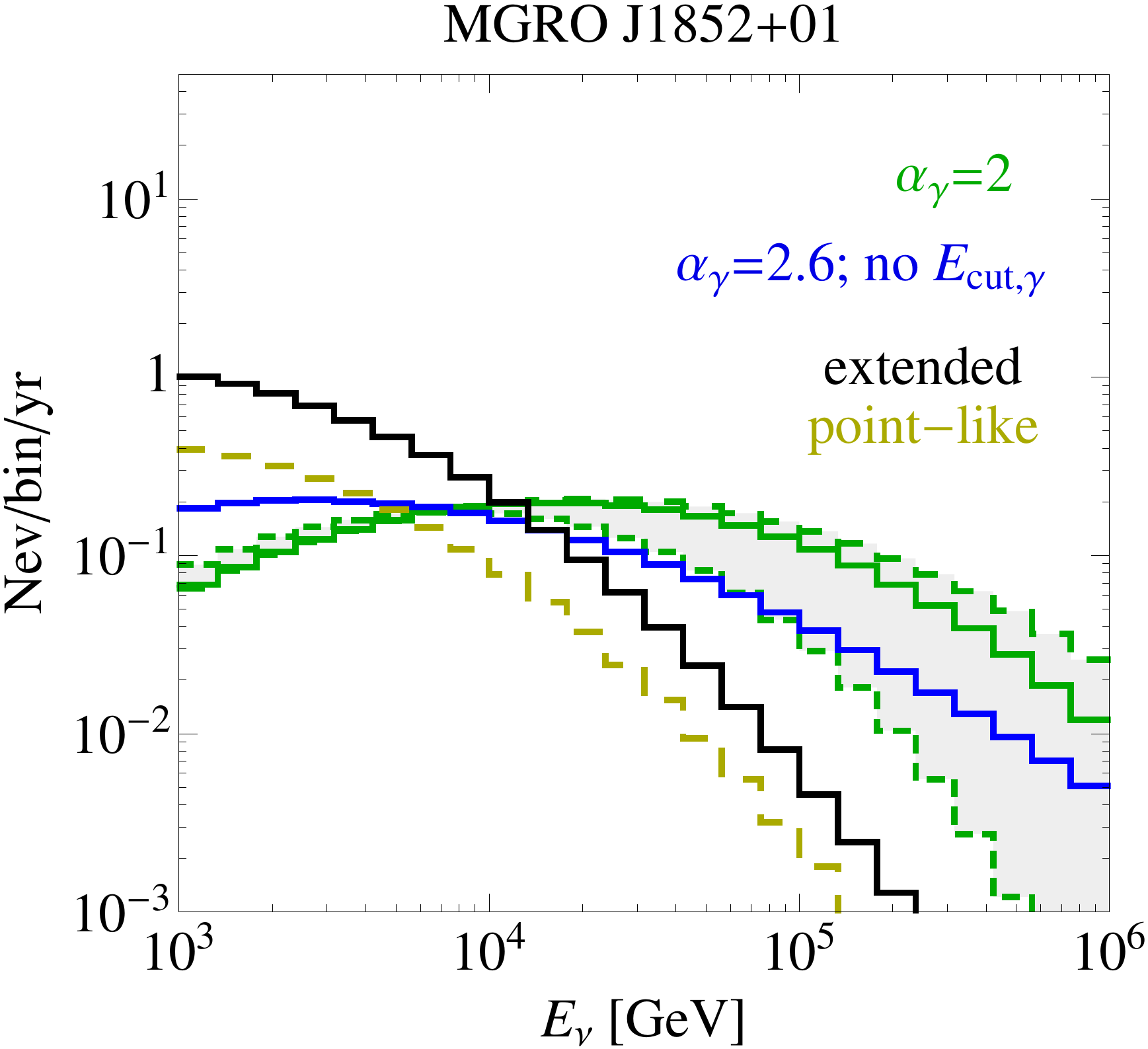} 
\end{tabular}
\caption{\label{fig:sources_spectra_forth} 
{\it \underline{Left panel:}} We show in blue the value on the 
flux reported by the Milagro collaboration~\cite{abdo}, which 
assumed an $E^{-2.6}$ spectrum. 
With green lines we show the spectra obtained considering 
$\alpha_\gamma=2$ and fixing the normalization to the best fit reported in 
Table~\ref{tab:sources_fit}, where we also allowed the cut-off energy to vary: 
$E_{\rm cut, \gamma} =30,~300, and~800$~TeV (short-dashed, solid, and long-dashed lines, in green). 
{\it \underline{Right panel:}} Number of events for the spectra reported with green and blue lines 
in the left panel. 
The gray band encodes the uncertainty on the cut-off energy. 
With the black (gold dashed) line, we show the background from atmospheric neutrinos for extended (point-like) sources. 
}
\end{figure}

\underline{\textbf{MGRO J1852+01:}} 
In the original Milagro survey, its statistical significance fell just below the statistical threshold to be a candidate source. With its recent observation by HAWC~\cite{hawc_gamma} MGRO J1852+01 becomes a plausible neutrino source candidate. 
The primary study of the six Milagro sources in Ref.~\cite{Halzen:2008zj} suggested that this source, 
due to its large flux, could considerably increase the probability of detecting neutrinos in IceCube in 
five years. 
The flux from a $3\times3$ degree region around MGRO J1852+01 is given by
$dN/dE = (5.7 \pm 1.5_{stat} \pm 1.9_{sys}) \times 10^{-14}~{\rm TeV^{-1}~cm^{-2}~s^{-1}}$ at the median detected 
energy of 12~TeV, assuming a differential source spectrum of $E^{-2.6}$~\cite{abdo}. In Table~\ref{tab:sources_ext} and Table~\ref{tab:sources_fit}, the information on this source is summarized, 
while in Fig.~\ref{fig:sources_spectra_forth} we show the best-fit spectrum from the Milagro collaboration. \\

\underline{\textbf{MGRO J2031+41:}} 
The flux from MGRO J2031+41 has been measured by Milagro~\cite{Abdo:2007ad,Abdo:2009ku,Abdo:2012jg}; the measurement cannot distinguish between a power law and a power law with cut-off. 
This is also the case for the ARGO-YBJ observations~\cite{Bartoli:2012tj}. The two experiments have comparable angular resolution. The flux measured by ARGO-YBJ ~\cite{Bartoli:2012tj} for this source is compatible with the one reported by Milagro, which extends to energy below 1~TeV. 

In general, ACT experiments report much smaller fluxes for this source. Indeed, measurements by MAGIC~\cite{Albert:2008yk}, HEGRA~\cite{Aharonian:2005ex}, and Whipple~\cite{Lang:2004bk} can account for just a few percent of the Milagro flux. The source has been recently studied by the VERITAS collaboration, which has reported a flux 
comparable to the one reported by MAGIC. In the current picture \cite{A.WeinsteinfortheVERITAS:2014iwa}, there are several sources contributing to the emission of MGRO J2031+41: the cocoon, the $\gamma$-Cygni SNR, VER J2019+407, and 
TeV J2032+4130. The latter has been detected by both VERITAS and MAGIC. 
In conclusion, a complete picture and understanding of this source is still not given. 
New data have been presented by the ARGO-YBJ detector in Ref.~\cite{Argo:2014tqa}, where the 
authors suggest identifying ARGO J2031+4157 as the TeV-energy counterpart of the 
Cygnus cocoon. For this reason, they report the best fit not only considering the ARGO-YBJ data 
but also including in their fit the Fermi-LAT data from the Cygnus cocoon. 
This results in a harder spectral index with significant consequences for the neutrino prediction. 
Since leptonic processes could contribute to the cocoon emission at the energies detected by Fermi-LAT, we might expect the purely hadronic 
component of MGRO J2031+41 to lie somewhere between the two fits obtained by the ARGO-YBJ collaboration. 
We report in Table~\ref{tab:sources_ext} the extension of MGRO J2031+41 as given by the ARGO-YBJ experiment, while in Table~\ref{tab:sources_fit} 
we show the fluxes obtained with and without the inclusion of the Fermi-LAT data in the fit. 
In Fig.~\ref{fig:sources_spectra_third}, we report the spectra for MGRO J2031+41 
from different experiments. Note that we do not report the measurements by HEGRA~\cite{Aharonian:2005ex} and 
Whipple~\cite{Lang:2004bk}, but these are in agreement with the MAGIC results. \\

\underline{\textbf{MGRO J2019+37:}} 
The flux of the source MGRO J2019+37 has been measured by Milagro, see 
Refs.~\cite{Abdo:2007ad,Abdo:2009ku,Abdo:2012jg}, reporting a power-law 
with energy cut-off as best fit. This source has not been detected by the ARGO-YBJ detector, which instead set 90\% C.L. upper bounds on the 
flux~\cite{Bartoli:2012tj}. Additionally, a limit on the flux at 115~TeV has been inferred through the CASA-MIA experiment~\cite{Beacom:2007yu}. 

The Milagro source MGRO J2019+37 has been recently detected by VERITAS. In Ref.~\cite{Aliu:2014rha}, the collaboration reported two sources in the region of 
MGRO J2019+37: the faint point-like source VER J2016+371 and the bright extended source VER J2019+368. 
This second source is likely to account for the bulk of the Milagro emission. The VERITAS collaboration reported a very low spectral index for this source on the order of 1.75, 
between 1--30~TeV. We list in Table~\ref{tab:sources_ext} the extension for MGRO J2019+37 as given by VERITAS and the Milagro 2012 release, and in Table~\ref{tab:sources_fit} the value of the flux reported by the VERITAS experiment. 
In Fig.~\ref{fig:sources_spectra_first}, we show the data for MGRO J2019+37 from different experiments. 

\vspace{0.4cm}

\begin{table}[!t]
\centering
\begin{tabular}{l || c | c | c }
\hline 
Source & Type & $\sigma_{\rm ext}$~(ACT) & $\sigma_{\rm ext}$~(EAS) \\ [1ex]\hline \hline
MGRO J1908+06 & UNID & & \\
$\hookrightarrow$ ARGO-YBJ & & & 0.49$^\circ \pm 0.22^\circ$~\cite{Bartoli:2012tj} \\
$\hookrightarrow$ HESS J1908+063 & & 0.34$^\circ$ $^{+0.04}_{-0.03}$~\cite{Aharonian:2009je} & \\
$\hookrightarrow$ VERITAS & & 0.44$^\circ \pm 0.02^\circ$~\cite{Aliu:2014xra} & \\ [1ex]\hline
MGRO J1852+01 & UNID &  & Milagro: $3^\circ \times 3^\circ$ search region~\cite{abdo}\\ [1ex]\hline
MGRO J2031+41 & UNID &  & \\
$\hookrightarrow$ ARGO J2031+4157 &  &  & 1.8$^\circ \pm 0.5^\circ$~\cite{Argo:2014tqa} \\[1ex] \hline
MGRO J2019+37 & PWN & & Milagro: 0.7$^\circ$~\cite{Abdo:2012jg} \\
$\hookrightarrow$ VER J2019+368 & & $\sim 0.35^\circ$~\cite{Aliu:2014rha} & \\[1ex] \hline 
\end{tabular}
\caption{Extensions of the sources as reported by different experiments. For the source MGRO J2031+41, 
we do not report the extension of the corresponding sources detected by ACT experiments, since the 
flux of these sources is much smaller than the one reported by the Milagro collaboration, 
see text for details. Note that the four sources have been recently detected by 
HAWC~\cite{Abeysekara:2015qba, hawc_gamma}.
}
\label{tab:sources_ext}
\end{table}

\begin{table}[!t]
\centering
\begin{tabular}{l || c }
\hline
Source & $E_\gamma^{\rm norm};~~~dN^{12}_\gamma/dE_\gamma {\rm \, at \,} E^{\rm norm}_\gamma;~~~\alpha_\gamma$ (ACT or EAS)\\
[1ex]\hline \hline
MGRO J1908+06 & \\ 
$\hookrightarrow$ HESS J1908+063 & 
1~TeV;~~~$4.14 \pm 0.32_{stat} \pm 0.83_{sys}$;~~~$2.10\pm0.07_{stat}\pm 0.2_{sys}$~\cite{Aharonian:2009je}\\
$\hookrightarrow$ VERITAS & 
1~TeV;~~~$4.23 \pm 0.41_{stat}\pm 0.85_{sys}$;~~~$2.20\pm 0.10_{stat} \pm 0.20_{sys}$~\cite{Aliu:2014xra}\\
\hline
MGRO J1852+01 & \\
$\hookrightarrow$ Milagro & 
12~TeV;~~~$(5.7 \pm 1.5_{stat} \pm 1.9_{sys})\times 10^{-2};$~~~2.6~\cite{abdo}\\ \hline
MGRO J2031+41 & \\ 
$\hookrightarrow$ ARGO J2031+4157 & 
w/o Fermi-LAT: \\
& 1~TeV;~~~$(2.5\pm 0.4)\times 10;~~~2.6\pm 0.3$~\cite{Argo:2014tqa}\\
& 
w Fermi-LAT: \\
& 0.1~TeV;~~~$(3.5\pm 0.3)\times 10^3;~~~2.16\pm 0.04$~\cite{Argo:2014tqa}\\
\hline 
MGRO J2019+37 & \\
$\hookrightarrow$ VER J2019+368 & 
5~TeV;~~~$(8.1 \pm 0.7_{stat} \pm 1.6_{sys})\times 10^{-2}$;~~~$1.75\pm0.08_{stat}\pm 0.2_{sys}$~\cite{Aliu:2014rha} \\ 
\hline
\end{tabular}
\caption{Flux in units of $10^{-12}~{\rm TeV}^{-1}~{\rm cm}^{-2}~{\rm s}^{-1}$ at a 
specific energy $E^{\rm norm}_\gamma$ and 
spectral index $\alpha_\gamma$ as recently reported by ACT or EAS experiments.}
\label{tab:sources_fit}
\end{table}

\begin{table}[!t]
\centering
\begin{tabular}{|l || c | c | c |}
\hline
Source & $\sigma_{\rm eff} (\textrm{point-like;~~extended})$ & Flux\\\hline\hline
MGRO~J1908+06 & $0.64^\circ;~~0.72^\circ$ & HESS~\cite{Aharonian:2009je} \\
MGRO~J1852+01 & $0.64^\circ;~~1.63^\circ$ & Milagro~\cite{abdo} \\
MGRO~J2031+41 & $-;~~1.91^\circ$ & ARGO-YBJ (+Fermi-LAT)~\cite{Argo:2014tqa}\\
MGRO~J2019+37 & $0.64^\circ;~~0.73^\circ$ & VERITAS~\cite{Aliu:2014rha}  \\\hline
\end{tabular}
\caption{
Angular opening and normalization for the flux considered in the analysis. The angle is 
defined as $\sigma_{\rm eff} \equiv \sqrt{\sigma_{\rm ext}^2 + \sigma_{\rm IC}^2}$, 
where $\sigma_{\rm ext}$ is 
the extension of the source reported by the collaboration (see Table~\ref{tab:sources_ext}), while 
$\sigma_{\rm IC} \equiv 1.6~\Delta \xi_{\rm IC}$, with 
$\Delta \xi_{\rm IC} =0.4^\circ$, is the IceCube angular resolution. For the source MGRO J1908+06, we 
will use the normalization given by HESS, compatible with the one reported by VERITAS; see Table~\ref{tab:sources_fit} 
for the specific values.}
\label{tab:r_bin}
\end{table}

\section{\label{sec:flux} Gamma rays and neutrino flux}
Perfoming a fit to the the gamma-ray flux using the parametrization 
\begin{equation}
\frac{dN_{\gamma}(E_\gamma)}{dE_\gamma}
=k_{\gamma}
 \left(\frac{E_\gamma}{\rm TeV}\right)^{-\alpha_{\gamma}} 
\exp\left(-\sqrt{\frac{E_\gamma}{E_{cut,\gamma}}}\right)\,,
\label{eq:Ngamma}
\end{equation}  
the neutrino fluxes at Earth can be described by 
the following expression~\cite{Kelner:2006tc,Kappes:2006fg}:
\begin{equation} 
\frac{dN_{\nu_\mu+\bar\nu_\mu}(E_\nu)}{dE_\nu}
=k_{\nu} \left(\frac{E_\nu}{\rm TeV}\right)^{-\alpha_ {\nu}}
\exp\left(-\sqrt{\frac{E_\nu}{E_{cut,\nu}}}\right)\,,
\label{eq:flux_nu}
\end{equation} where
\begin{eqnarray}
&& k_\nu=(0.694-0.16 \alpha_\gamma) k_\gamma \nonumber\,, \\ &&
  \alpha_\nu= \alpha_\gamma\,, \nonumber \\ && 
E_{cut,\nu}=0.59  E_{cut,\gamma}\,.
\label{eq:params_nu}
\end{eqnarray}

The number of throughgoing muon neutrinos from a source at zenith angle $\theta_Z$ is given by Ref.~\cite{Halzen:2008zj}:
\begin{eqnarray}
N_{ev}= t\, \int_{E_\nu^{\rm th}} dE_\nu ~\frac{dN_\nu(E_\nu)}{dE_\nu} \times A_\nu^{\rm eff}(E_\nu,\theta_Z)\,, 
\label{eq:nevmus}
\end{eqnarray}
where we have summed over neutrino and antineutrino contributions. We will use the IceCube neutrino effective area reported in Ref.~\cite{Aartsen:2014cva}.

\section{\label{sec:res} Results}
Based on the updated information from gamma-ray experiments described in the previous section, we revisit the prospects for 
observing neutrinos from these sources with IceCube, using the effective area for the 86-string detector configuration~\cite{Aartsen:2014cva}. 
This study updates a previous study of three of the sources \cite{Gonzalez-Garcia:2013iha} using Milagro~\cite{Abdo:2012jg,Smith:2010yn} 
and ARGO-YJB~(2012) measurements~\cite{Bartoli:2012tj}. 
For related studies of the neutrino emission from Milagro sources, see also 
Refs.~\cite{Beacom:2007yu,Halzen:2008zj,Kappes:2009zza,Halzen:2007ah,Vissani:2011ea,Vissani:2011vg,Tchernin:2013wfa}. 

The new information from gamma-ray experiments turns out to be important for a better parametrization of the flux of the gamma-ray sources. 
The uncertainties in the normalization and spectrum of the sources can result in important variations in the prediction of the neutrino fluxes. In this context, using updated data is important to make more reliable predictions and more appropriate interpretations of potential IceCube observations.

\begin{figure}[!t]
\includegraphics[width=0.85\textwidth,height=8cm]{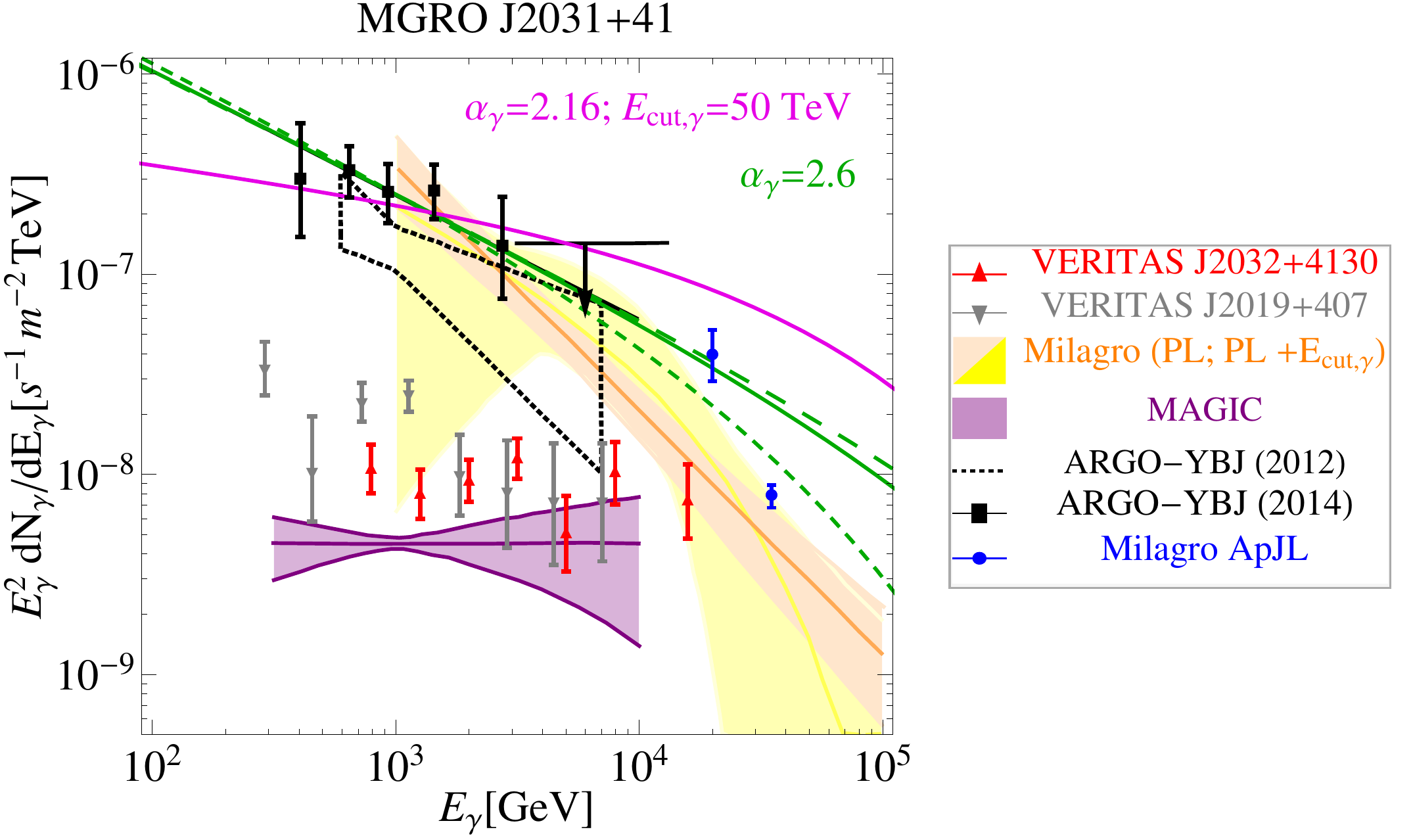} \vspace{0.5cm} \\
\hspace{-1cm}
\includegraphics[width=0.48\textwidth]{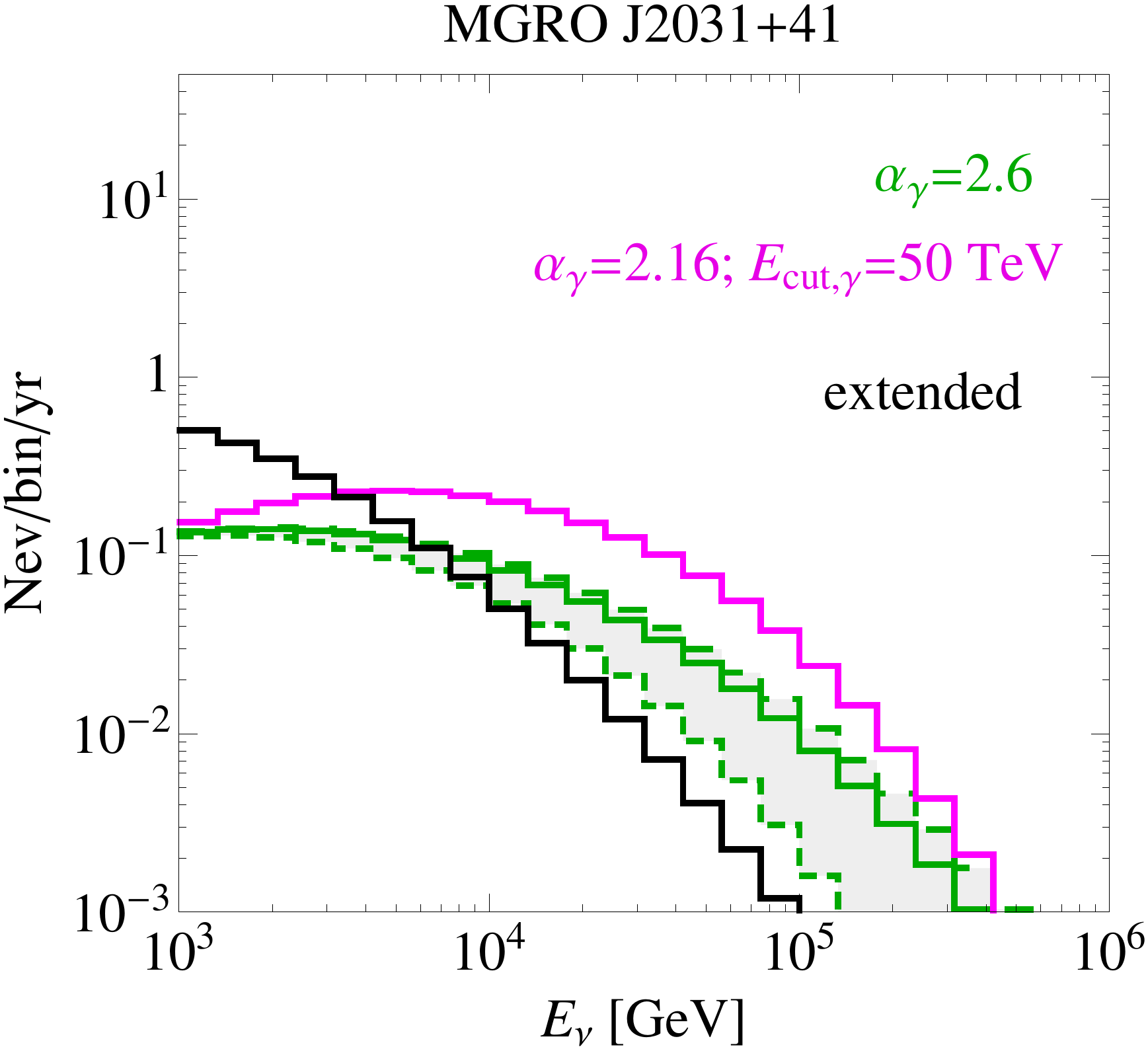} 
\includegraphics[width=0.48\textwidth]{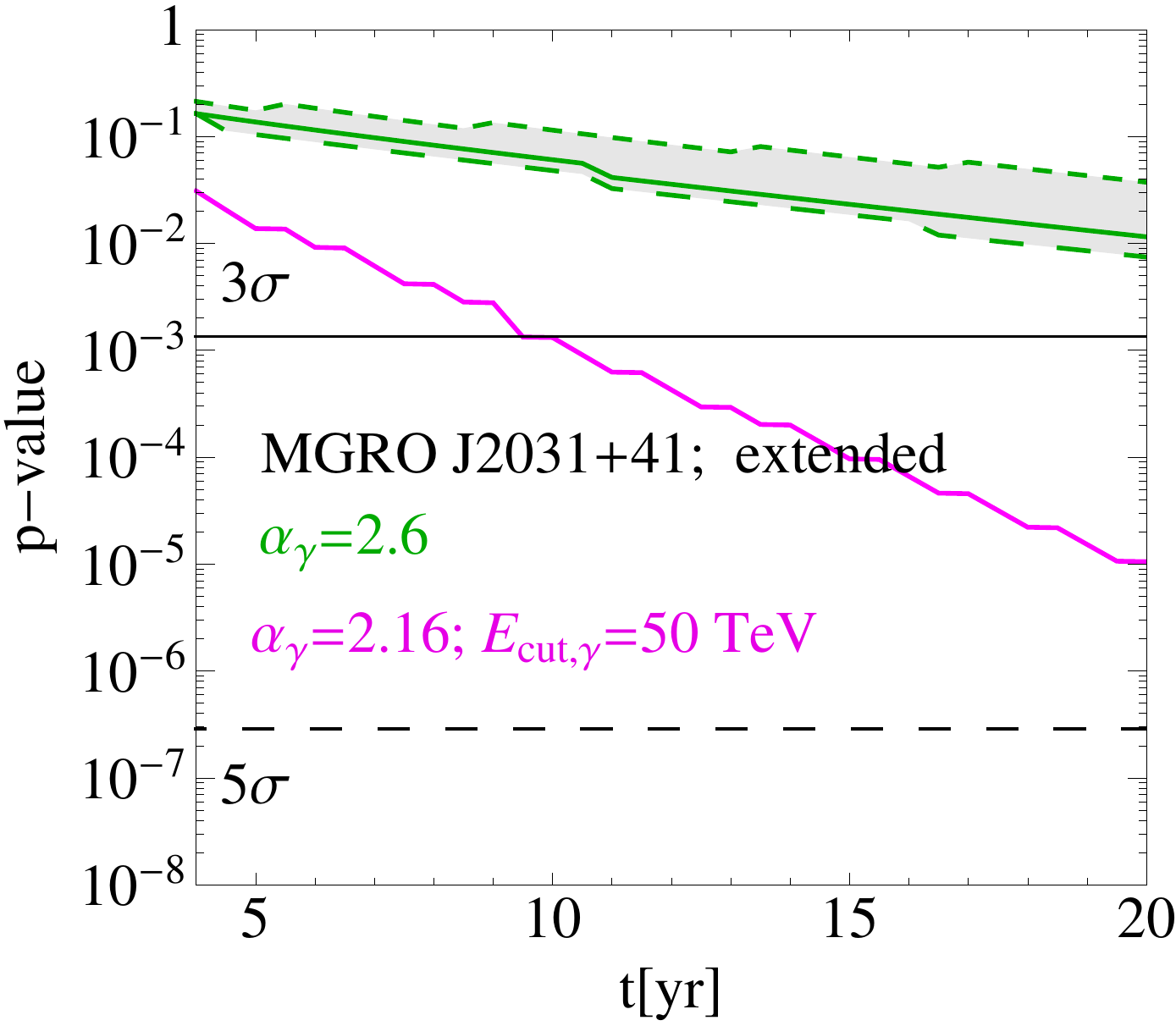} 
\caption{\label{fig:sources_spectra_third} 
{\it \underline{Upper panel:}} 
The black points show the data reported by ARGO-YBJ in Ref.~\cite{Argo:2014tqa}, while 
the dotted region is the one reported in Ref.~\cite{Bartoli:2012tj}. 
The previous flux measurements by Milagro are 
shown in blue~\cite{Abdo:2007ad,Abdo:2009ku}, while the orange/yellow 
area denotes the the power-law model/the power-law model with 
cut-off as reported in Ref.~\cite{Abdo:2012jg} by Milagro. 
With the purple band we report the measurements by MAGIC~\cite{Albert:2008yk}. 
We report in red and grey the results from the VERITAS detector~\cite{A.WeinsteinfortheVERITAS:2014iwa}. 
With green/magenta lines we show the spectra obtained fixing the parameters to the best fit reported in 
Table~\ref{tab:sources_fit} for the case without/with Fermi data. In the case without Fermi data, 
we also allowed the cut-off energy to vary: 
$E_{\rm cut, \gamma} =30,~300, and~800$~TeV (short-dashed, solid, and long-dashed lines, in green. 
{\it \underline{Lower panel, left:}} Number of events for the spectra reported with green and magenta lines 
in the upper panel. The gray band encodes the uncertainty on the cut-off energy. 
With black lines, we show the background from atmospheric neutrinos. 
{\it \underline{Lower panel, right:}} p-values as a function of time, from 4 years to 
20 years.
}
\end{figure}

\begin{figure}[!t]
\includegraphics[width=0.48\textwidth]{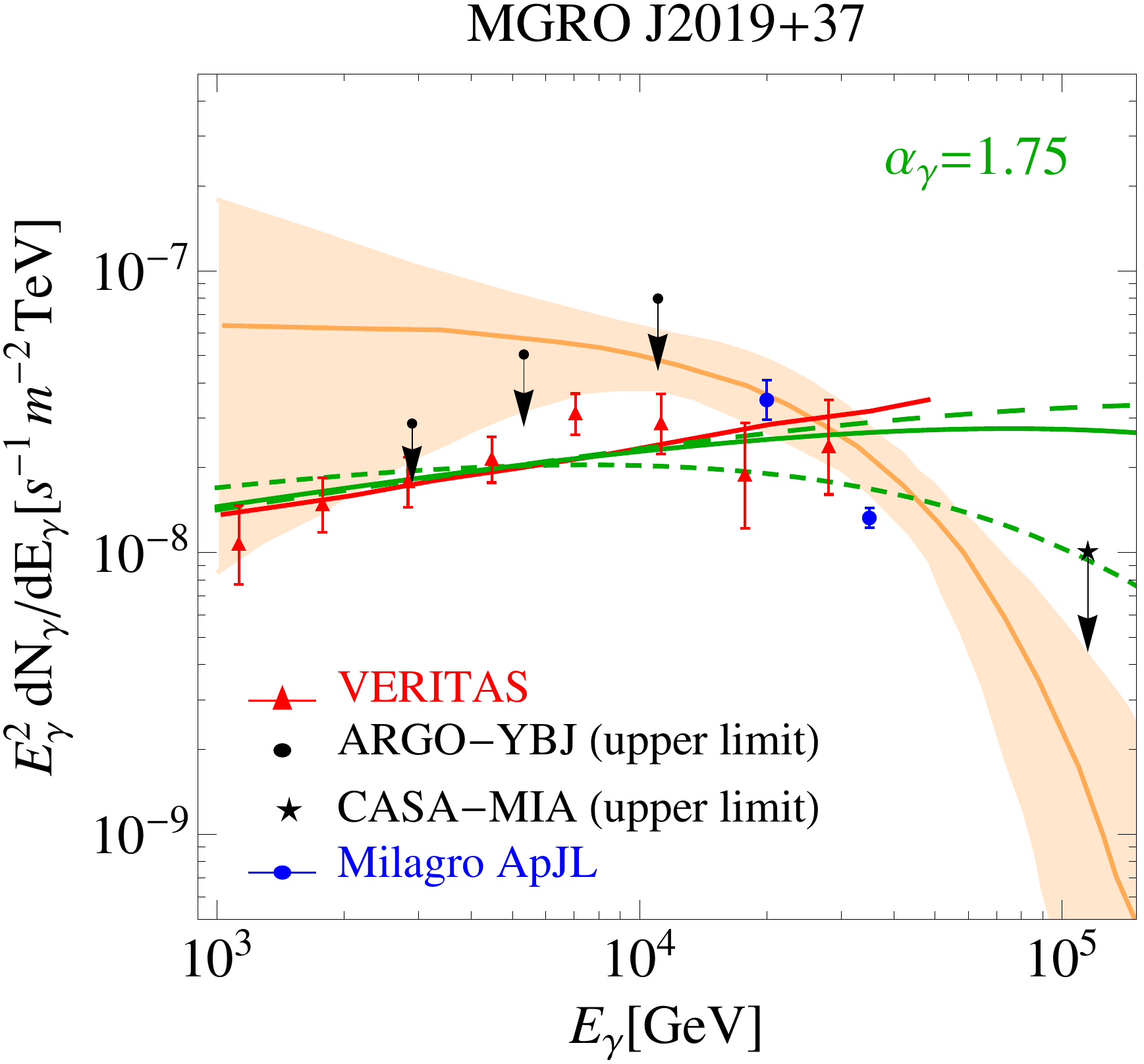} \\
\begin{tabular}{rl}
\includegraphics[width=0.48\textwidth]{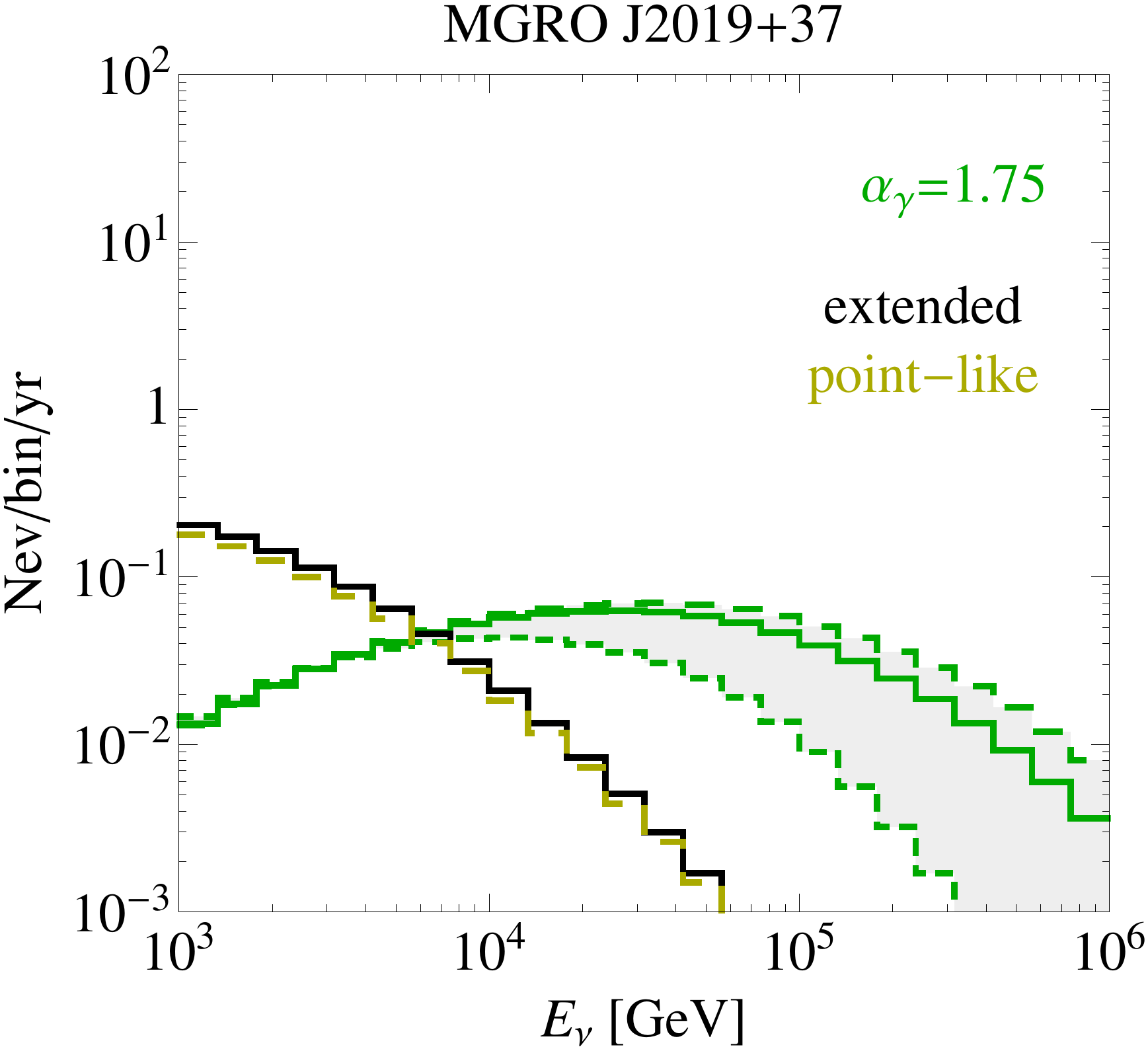} &
\includegraphics[width=0.48\textwidth]{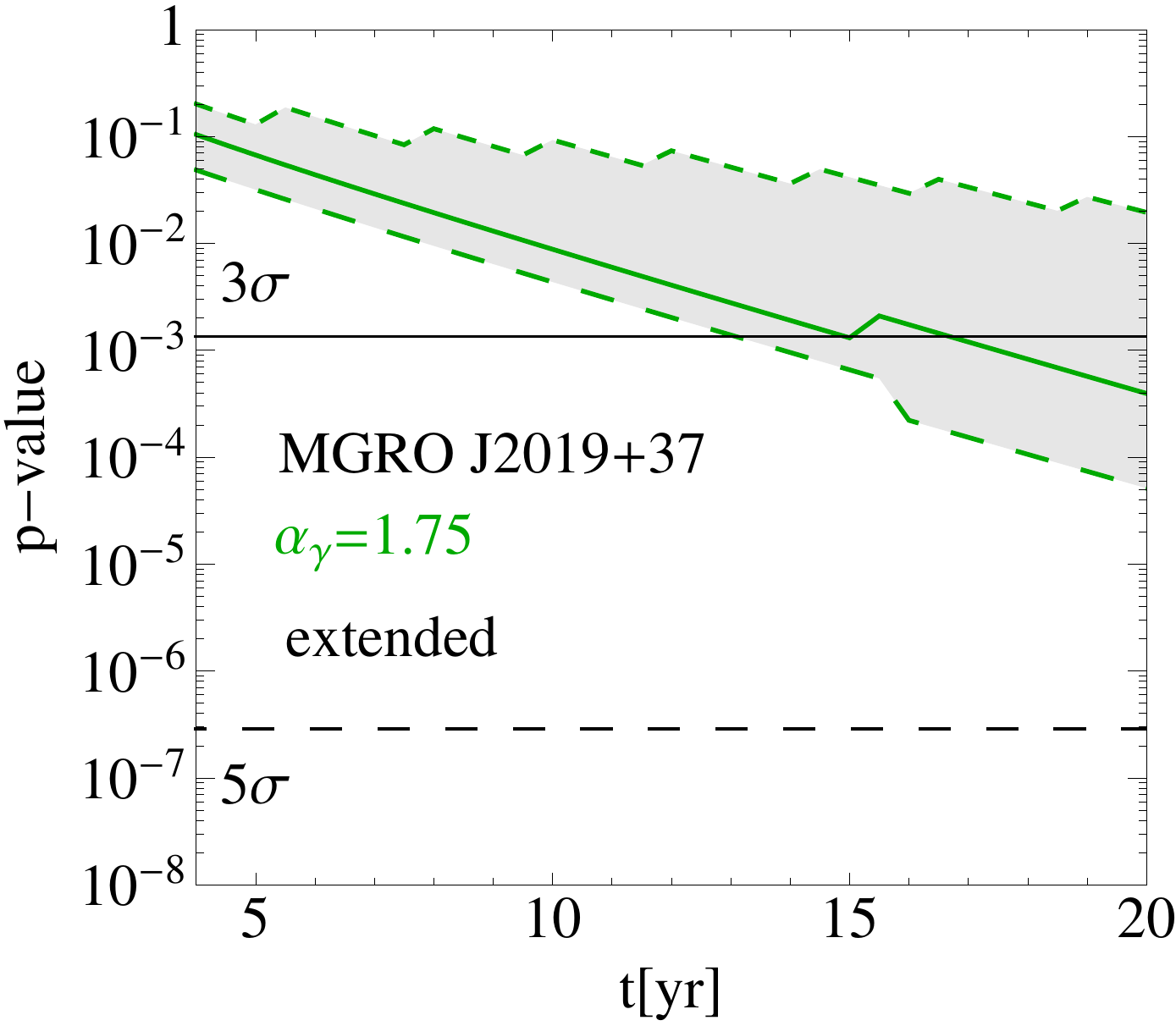} \\
\end{tabular}
 
\caption{\label{fig:sources_spectra_first} 
{\it \underline{Upper panel:}} 
With red points, we report the VERITAS data~\cite{Aliu:2014xra}. 
With blue lines, we
report the previous flux measurements by Milagro~\cite{Abdo:2007ad,Abdo:2009ku}, while 
the continuous orange
line and the shaded orange area
represent the best fit and $1\sigma$ band~\cite{Abdo:2012jg} as reported by Milagro. 
The 90\% C.L. upper limits from
ARGO-YBJ are shown in black~\cite{Bartoli:2012tj}, and the inferred CASA-MIA 
bound~\cite{Beacom:2007yu} is shown with a black star. 
With green lines we show the spectra obtained fixing the parameters to the best fit reported in 
Table~\ref{tab:sources_fit}, where we also allowed the cut-off energy to vary: 
$E_{\rm cut, \gamma} =30,~300, and~800$~TeV (short-dashed, solid, and long-dashed lines, in green).  
{\it \underline{Lower panel, left:}} Number of events for the spectra reported with green and magenta lines 
in the left panel. The gray band encodes the uncertainty on the cut-off energy. 
With the black (gold dashed) line, we show the background from atmospheric neutrinos for extended (point-like) sources. 
{\it \underline{Lower panel, right:}} p-values as a function of time, from 4 years to 
20 years.
}
\end{figure}

After calculating the neutrino flux, we compute the number of through-going muon neutrinos in IceCube. These have been produced inside or below the detector by neutrinos that have traversed the Earth. Any background of cosmic ray muons has thus been filtered out and only atmospheric neutrinos remain as a background for the northern hemisphere sources in a detector located at the South Pole. For each source, we fix the flux normalization to the best-fit values listed in Table~\ref{tab:sources_fit}. 
The expected number of muon neutrinos per energy bin are shown in 
Figs.~\ref{fig:sources_spectra_second},~\ref{fig:sources_spectra_forth},~\ref{fig:sources_spectra_third} and~\ref{fig:sources_spectra_first} for the 
four sources considered in the analysis. For MGRO~J1908+06, we have fixed $\alpha_\gamma =2$, consistent with the 
value reported by HESS, and we have varied the cut-off energy from 30~TeV up to 800~TeV. For MGRO~J1852+01, 
besides assuming $\alpha_\gamma =2$, we have also considered $\alpha_\gamma =2.6$ because this is the spectrum assumed by the Milagro collaboration. 
For MGRO~J2031+41, we have considered the best-fit values for $\alpha_\gamma$ provided by the ARGO-YBJ collaboration, 
considering also the case in which the Fermi-LAT data have been added to the fit. Finally, for MGRO~J2019+37 
we have considered the case of $\alpha_\gamma \sim 1.75$, the best-fit value reported by the VERITAS collaboration. 

To calculate the number of background atmospheric neutrino events, we have integrated the atmospheric flux \cite{Honda:2011nf} over an opening 
angle $\Omega=\pi \sigma_{\rm eff}^2$ around 
the direction of the source, where the angle $\sigma_{\rm eff}=\sqrt{\sigma_{\rm ext}^2 + \sigma_{\rm IC}^2}$. The values of $\sigma_{ext}$, the physical extension of the source, are reported in 
Table~\ref{tab:sources_ext}, while $\sigma_{\rm IC} \equiv 1.6~\Delta \xi_{\rm IC}$. The angular resolution of the IceCube detector $\Delta \xi_{\rm IC}$ is $0.4^\circ$ at the energies relevant for this analysis \cite{Aartsen:2015rwa}. This angular radius correspond to a solid angle that contains roughly 72\% of the signal events from the source; see also Ref.~\cite{Alexandreas:1992ek} for a discussion. 
The difference in the neutrino fluxes assuming an extended and point source is important as can be seen from Table~\ref{tab:r_bin}, with the biggest difference between these two assumptions for MGRO~J1852+01. 

We have subsequently estimated the statistical significance for observing the sources using the analytic expression ~\cite{ATLAS:2011tau}: 
\begin{equation}
p_{\rm value}=\frac{1}{2}\left[ 1-{\rm{erf}} 
\left( \sqrt{q_0^{obs}/2} \right) \right]\,,
\end{equation}
where $q_0^{obs}$ is defined as 
\begin{equation}
q_0^{obs} \equiv -2 \ln \mathcal{L}_{b,D}= 2  \left( Y_{b} - N_{D} + 
N_{D} \ln \left( \frac{N_{D}}{Y_{b}}\right)\right)\,.
\end{equation}
Here, $Y_{b}$ is the theoretical expectation for the background hypothesis, while 
$N_{D}$ is the estimated signal generated as the median of events Poisson-distributed around 
the signal plus background. 
We have considered the total number of events (not binned in energy) to have a closer prediction to what is done in the IceCube point-source searches ~\cite{Aartsen:2014cva}. 

\begin{figure}[!t]
\centering
\begin{tabular}{rl}
\includegraphics[width=0.48\textwidth]{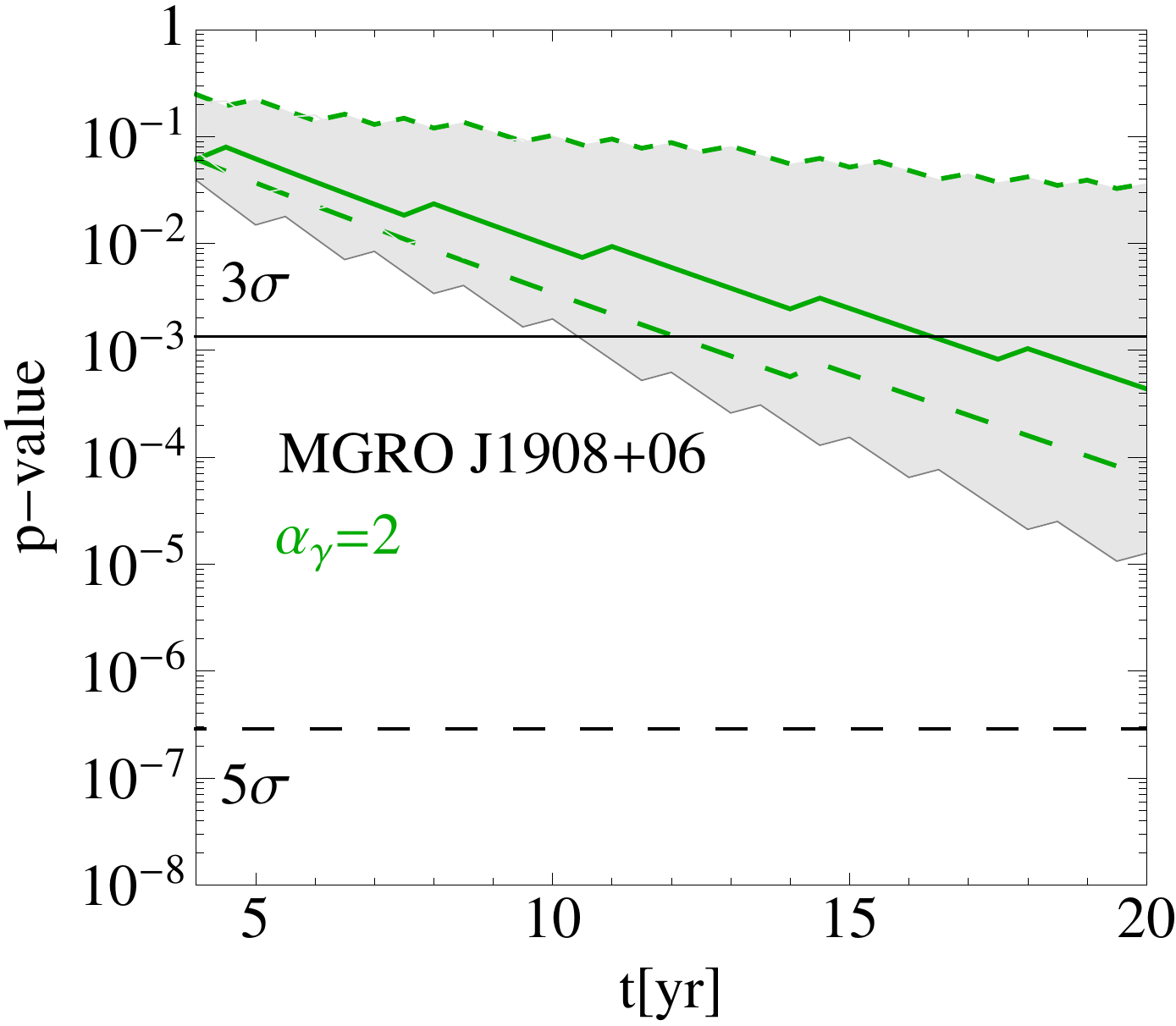} &
\includegraphics[width=0.48\textwidth]{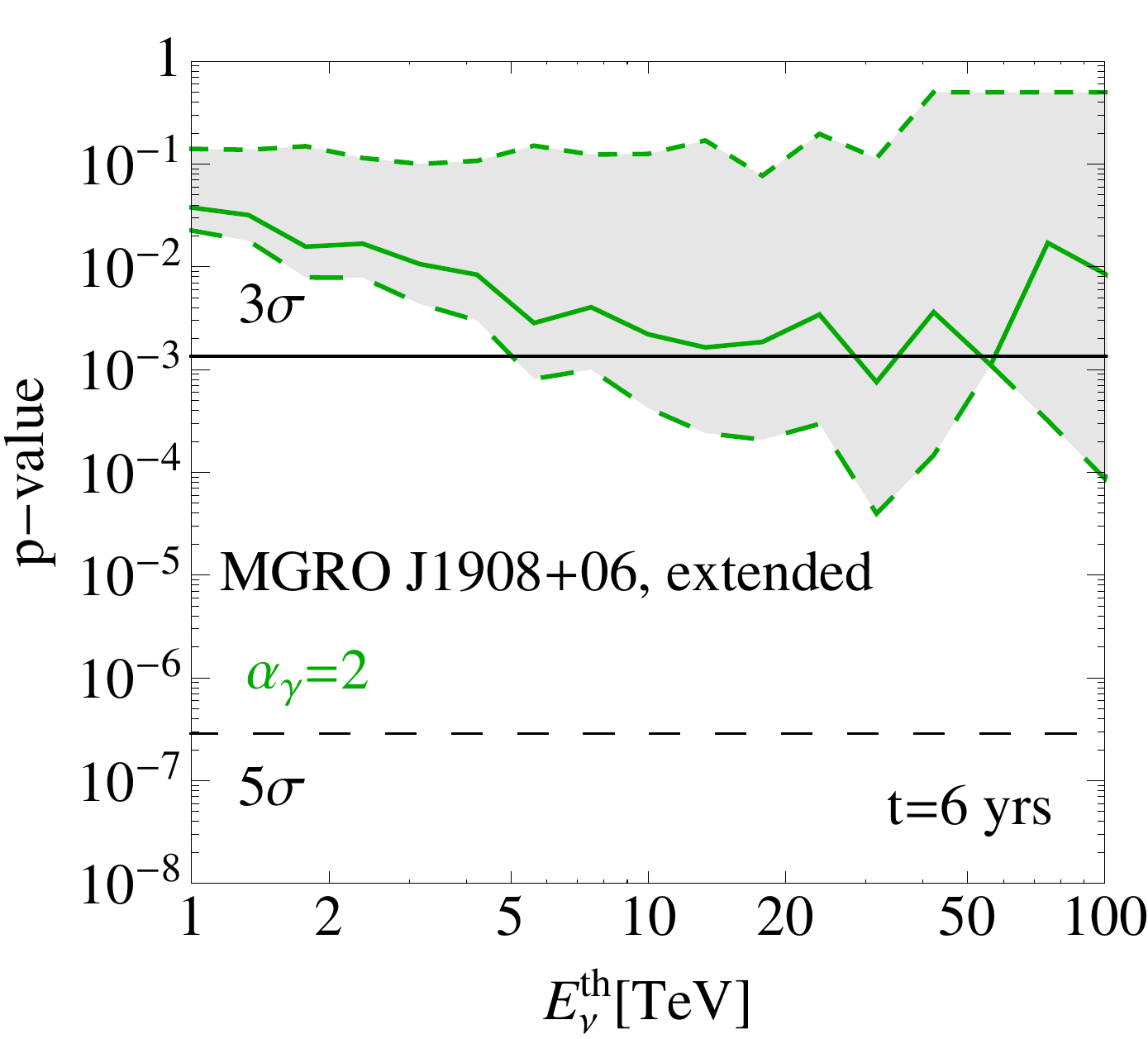} 
\end{tabular}
\caption{\label{fig:sources_pvalues_second} 
{\it \underline{Left panel:}}
p-values as a function of time, from 4 years to 
20 years. The 
spectra have been fixed, as shown in Fig.~\ref{fig:sources_spectra_second}. 
The gray band encodes the uncertainty due to different values of 
$E_{cut,\gamma}$, and morphology, see Table.~\ref{tab:r_bin}. For the green lines we
have considered the case of extended source. 
{\it \underline{Right panel:}} Dependence of the p-value on the 
energy threshold $E_{\nu}^{th}$. }
\end{figure}

In Fig.~\ref{fig:sources_pvalues_second}, we show the results for MGRO J1908+06. For this source, 
recent ACT data have reported a spectral index $\alpha_\gamma$ that is compatible with $\sim 2$. Despite the fact that the ACTs' normalization is smaller than the one previously reported by Milagro, the hard spectral index makes the 
source an interesting candidate for neutrino detection. For this reason, we also 
estimated how the p-value depends on the threshold energy that can be reached in a realistic analysis. We find that a $3\sigma$ 
discovery is possible in six years, if an energy threshold of about 5~TeV can be reached in the analysis, and that the spectrum extends to $E_{{\rm cut}, \gamma}$ of 800~TeV. For the more conservative case that $E_{{\rm cut}, \gamma} \sim$~300~TeV, 
as expected for galactic sources able to explain the cosmic-ray spectrum up to the knee, then an energy 
threshold of about 10~TeV would be required. Obtaining a 3$\sigma$ discovery at a specific 
energy threshold will indicate a particular value of the cut-off energy $E_{{\rm cut},\gamma}$. 

\begin{figure}[!t]
\centering
\begin{tabular}{rl}
\includegraphics[width=0.48\textwidth]{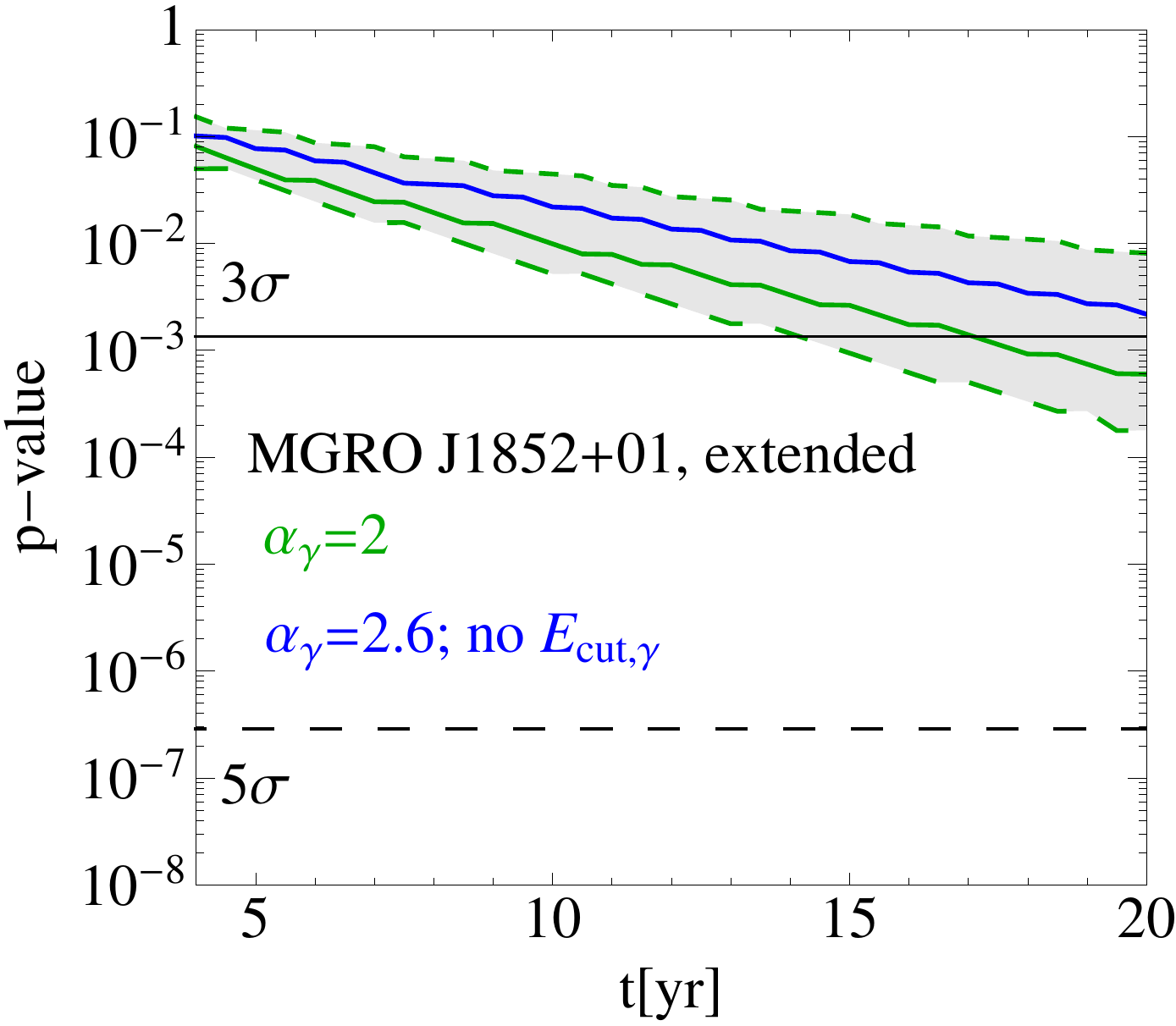} &
\includegraphics[width=0.48\textwidth]{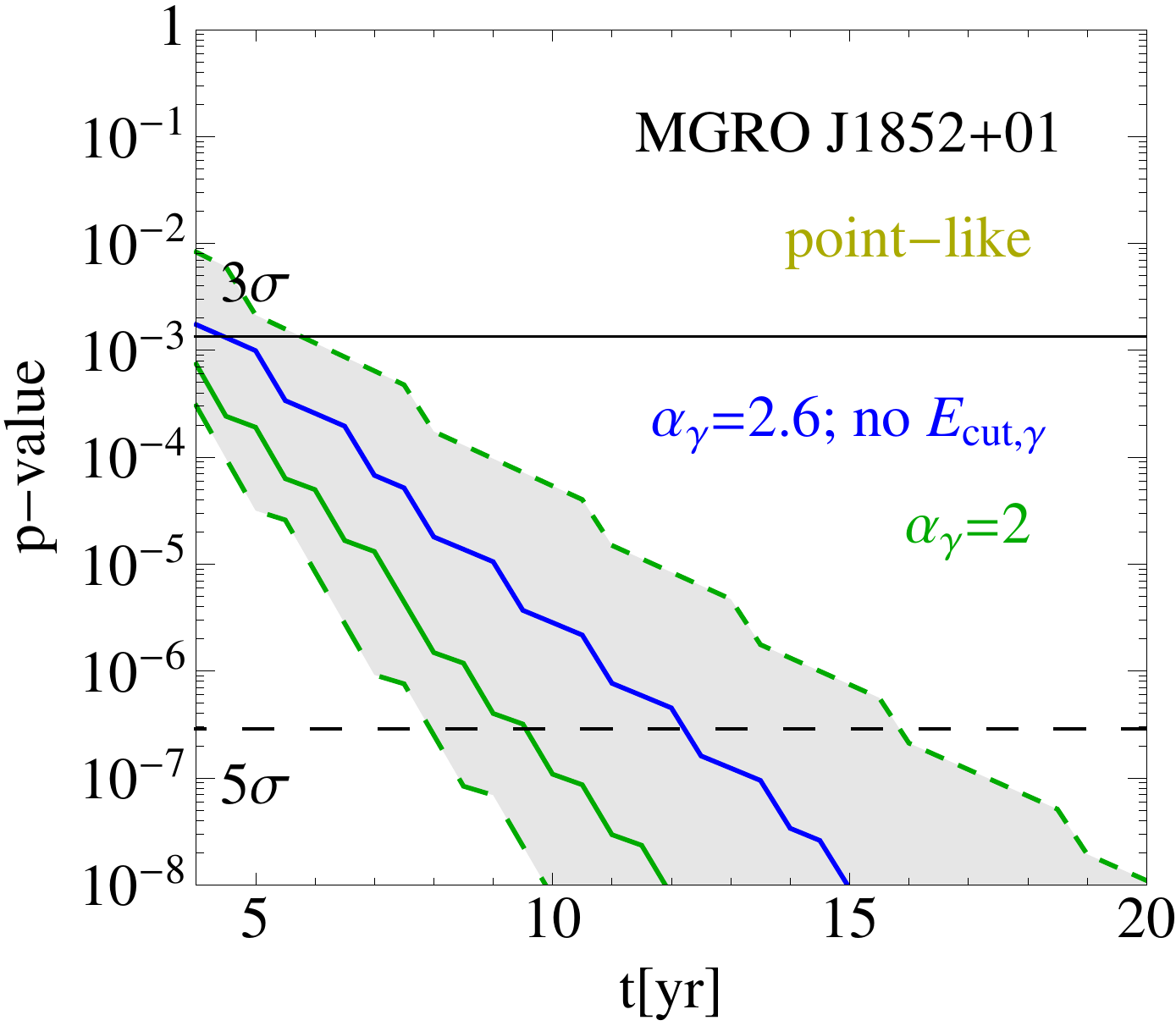} 
\end{tabular}
\caption{\label{fig:sources_pvalues_forth} 
{\it \underline{Left panel:}} 
p-values as a function of time, from 4 years to 
20 years. The 
spectra have been fixed, as shown in Fig.~\ref{fig:sources_spectra_forth}. 
The gray band encodes the uncertainty due to different values of 
$E_{cut,\gamma}$. We assume the source to be extended. 
{\it \underline{Right panel:}} We assume the source to be point-like.  
}
\end{figure}

In Fig.~\ref{fig:sources_pvalues_forth}, we show the statistical significance for MGRO J1852+01. For this 
source, due to the lack of data, not only is the spectral index poorly known but also the 
morphology of the source, whether extended or point-like, is uncertain. For the point-like 
hypothesis, a $3\sigma$ discovery can be reached in six years, independently of the energy cut-off $E_{{\rm cut},\gamma}$ and 
spectral index $\alpha_\gamma$ of 
the source, while more than 10 years are required if the source is extended. 

In Fig.~\ref{fig:sources_spectra_third}, we show the p-value for MGRO J2031+41. As explained in detail in 
the previous section, the origin of the gamma-ray emission from this source is not understood. Using the best fit obtained by the ARGO-YBJ collaboration in conjunction with the Fermi-LAT data, we find that a 3$\sigma$ discovery is possible with 10 years of IceCube data. If this is indeed realized, the IceCube data not only would point towards a hadronic emission at TeV energies for MGRO J2031+41 but would help clarify the origin of the gamma-ray emission from the cocoon. 

In Fig.~\ref{fig:sources_spectra_first}, we show the statistical significance for MGRO J2019+37. A detection of neutrinos from this source would be extremely interesting since it might point towards the mechanism described in Ref.~\cite{Lemoine:2014ala} for neutrino production in PWNs. 
For this source, we expect to obtain a 3$\sigma$ discovery in roughly 15 years. Future data from HAWC on 
the spectrum of this source are important to confirm the hard spectral index, on the order of $\alpha_\gamma \sim 1.75$, reported by the VERITAS collaboration. 

The Milagro collaboration has presented results on the energy spectrum of these sources obtained by unfolding of the data ~\cite{Abdo:2012jg}. It is obviously important for other experiments to confirm the presence of a low-energy cut-off that they consistently find in the analysis of every source. In this context, the constraints that IceCube can set in the plane ($\alpha_\gamma$,$E_{cut,\gamma}$) with future data are important and complementary. We have therefore estimated the constraints set on $\alpha_\gamma$ and $E_{\rm{cut}, \gamma}$ in 
the absence of a signal after 15 years of exposure with the complete 86-string IceCube detector. 
We have integrated the number of events from 1~TeV to 1~PeV in neutrino energy $E_{\nu}$ and 
defined the confidence level, C.L., as in Refs.~\cite{Junk:1999kv,Read:2000ru,ATLAS:2011tau,Beringer:1900zz}: 
\begin{equation}
C.L. = \frac{P_{(s+b)}}{1-P_b}\,.
\end{equation}
with $P_{(s+b)}$ and $P_b$ the p-values for the signal plus
background and background-only hypothesis of the data, respectively; see Ref.~\cite{Gonzalez-Garcia:2013iha} for details. 

\begin{figure}[!t]
\centering
\begin{tabular}{rl} 
\includegraphics[width=0.48\textwidth]{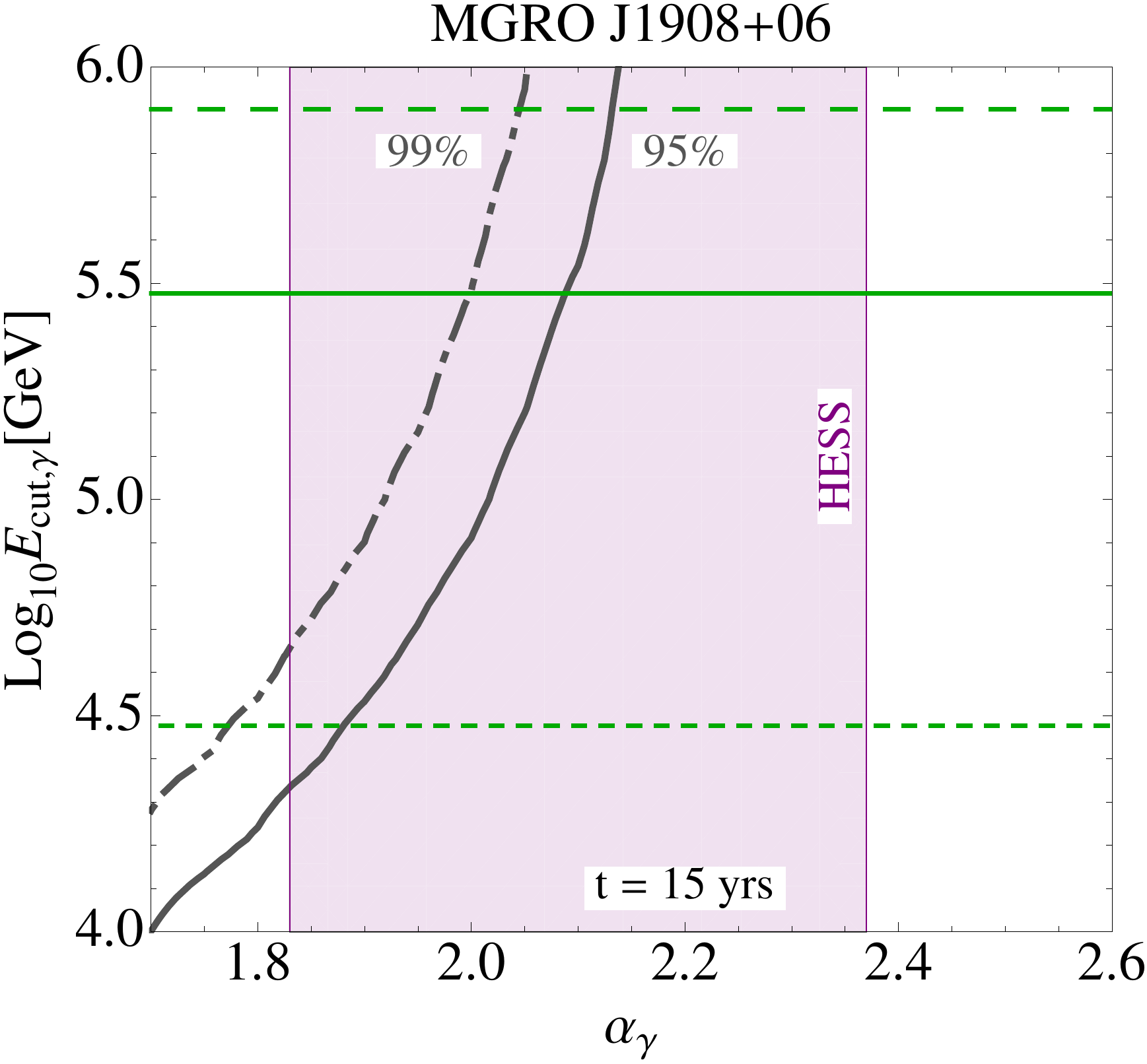} & 
\includegraphics[width=0.48\textwidth]{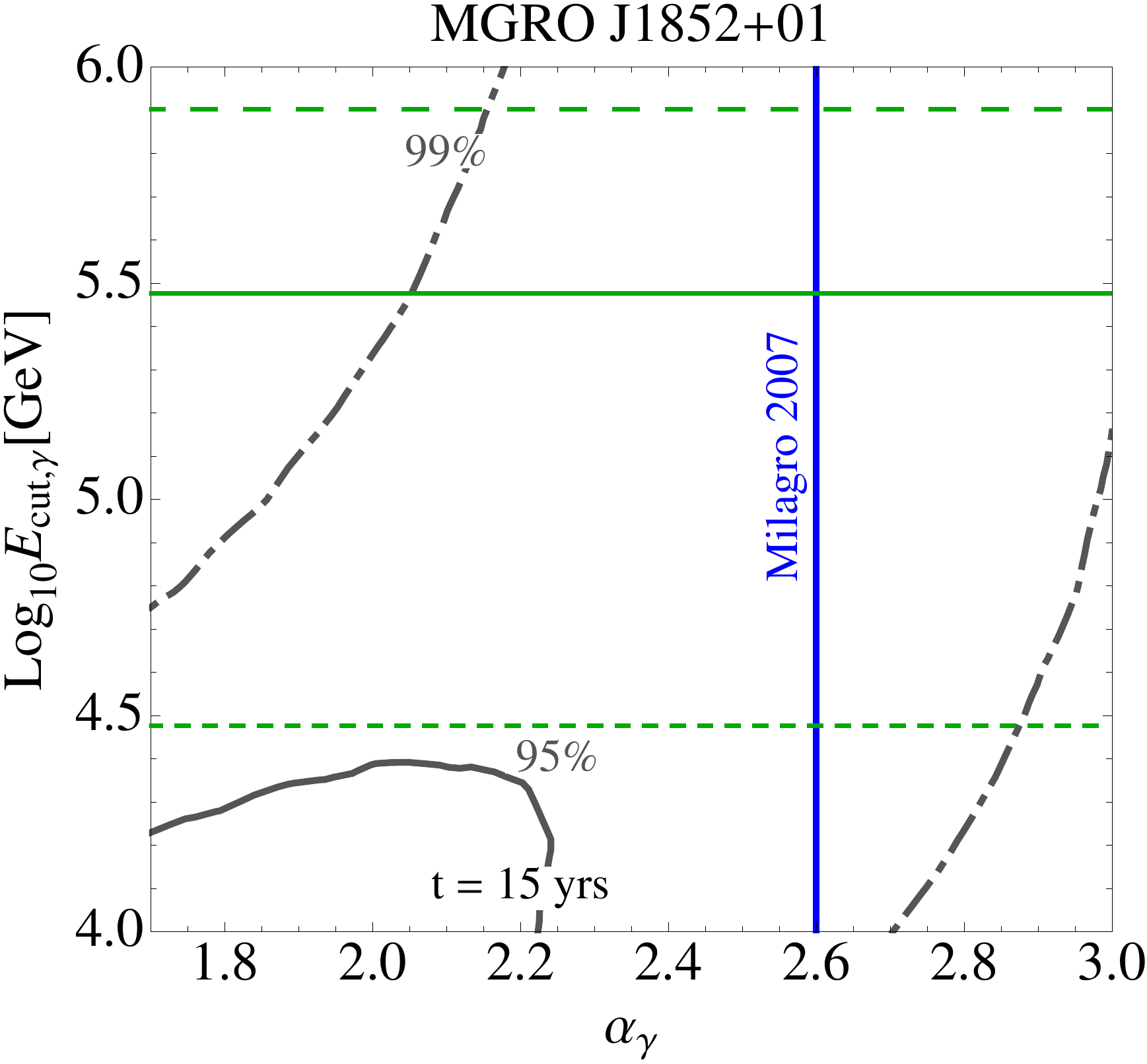} \vspace{0.5cm} \\
\includegraphics[width=0.48\textwidth]{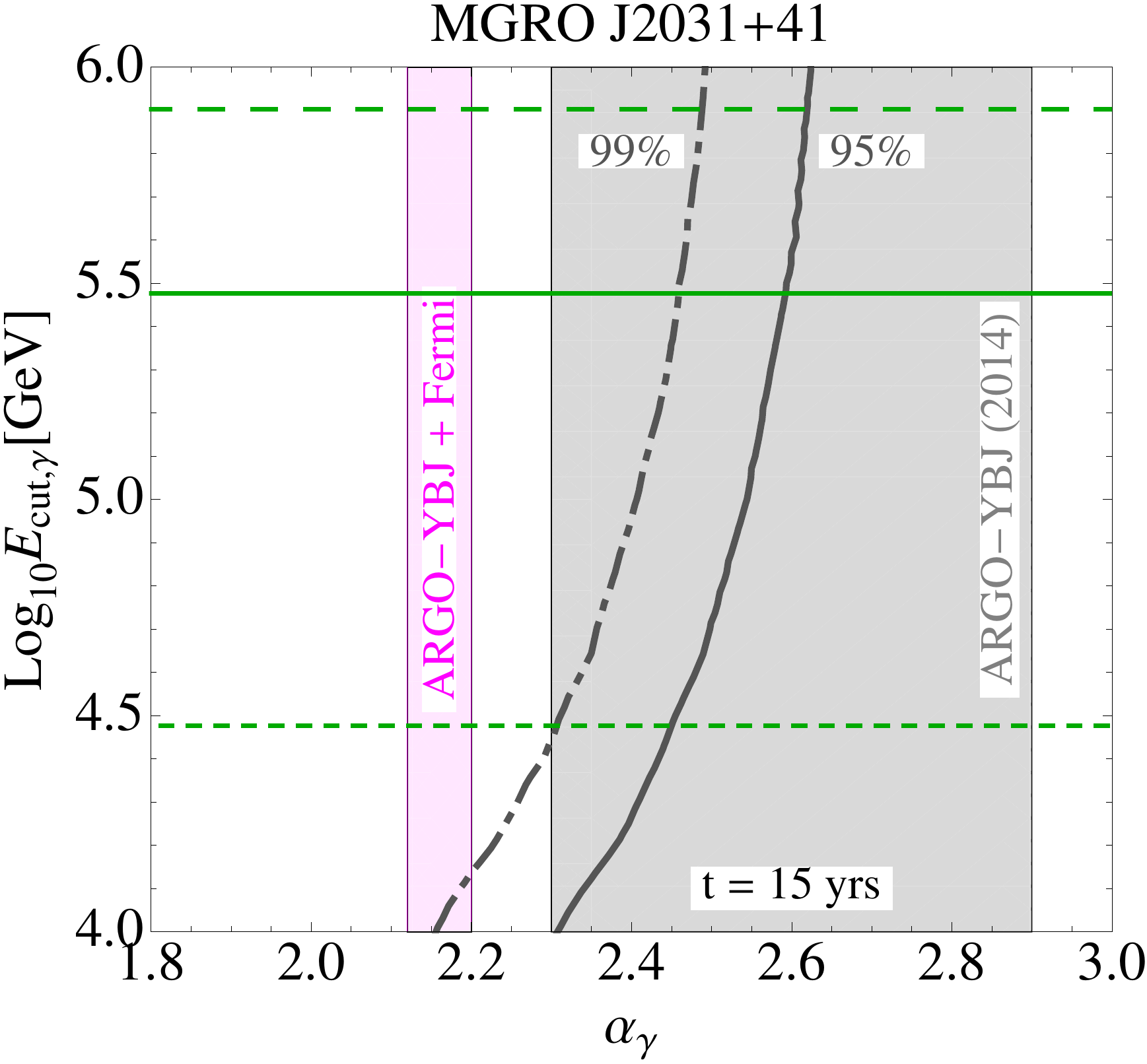} & 
\includegraphics[width=0.48\textwidth]{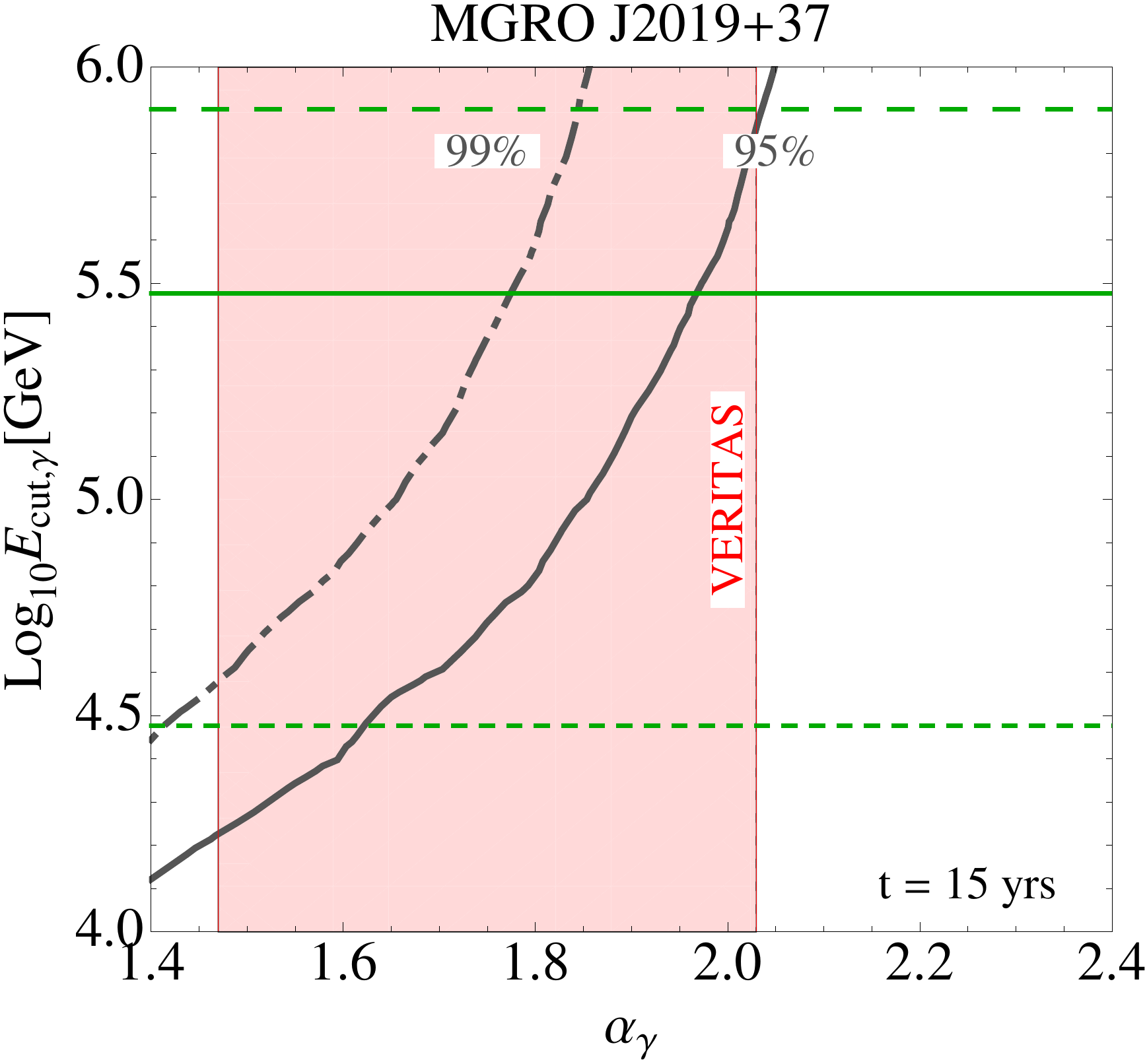}
\end{tabular}
\caption{\label{fig:sources_fixedNorm_2} 
{\it \underline{Upper panel:}} 
Values of $\alpha_\gamma$ and $E_{cut, \gamma}$ excluded at 95\% (solid) and 99\%~C.L. (dot-dashed) with 15 years of 
IceCube running with its 86-string configuration. 
The normalization has been fixed to the best fit reported by HESS~\cite{Aharonian:2009je} (left) 
and Milagro~2007 (right). We have assumed extended sources. With horizontal lines we denote the values 
$E_{\rm cut, \gamma} =30,~300, and~800$~TeV (short-dashed, solid, and long-dashed lines, in green). 
The purple region (left) denotes the values of $\alpha_\gamma$ reported by HESS. 
The blue line (right) denotes the value of $\alpha_\gamma$ considered by Milagro. 
{\it \underline{Lower panel:}} 
The normalization has been fixed to the best fit reported by ARGO-YBJ without Fermi-LAT~\cite{Bartoli:2012tj} (left) 
and VERITAS~\cite{Aliu:2014rha} (right). We have assumed extended sources. 
The gray/magenta region (left) denotes the values of $\alpha_\gamma$ reported by ARGO-YBJ without/with 
Fermi data. 
The red region (right) denotes the values of $\alpha_\gamma$ reported by VERITAS. 
}
\end{figure}

The results for the expected C.L. are presented in Fig.~\ref{fig:sources_fixedNorm_2} for the 
four sources considered in this paper and for a running time of $t$=15 years. 
We have fixed the normalization to the best fit reported in Table~\ref{tab:sources_fit}, while we have 
varied the values of the spectral index $\alpha_\gamma$ and the cut-off energy $E_{cut,\gamma}$.
As shown in the figure, for MGRO J1908+06, IceCube is able to 
constrain a major part of the values for $\alpha_\gamma$ reported by the HESS detector. 
In particular, for a spectral index as hard as $\alpha_\gamma \sim 2$, values of $E_{{\rm cut}, \gamma}$ 
greater than 100~TeV could be excluded at 95\%~C.L. For MGRO J1852+01, IceCube will exclude all the parameter space with 
$E_{{\rm cut}, \gamma}$ greater than 30~TeV at 95\%~C.L. 
For the source MGRO J2031+41, the allowed region of $\alpha_\gamma$ obtained considering ARGO-YBJ plus Fermi-LAT data 
will be excluded at 99\%~C.L., independently of the value of $E_{{\rm cut}, \gamma}$. 
Finally, for MGRO J2019+37, considering the standard value of $E_{{\rm cut},\gamma}$ of 300~TeV, 
hard values of the spectral index with $\alpha_\gamma < 2$ will be excluded at 95\%~C.L. 

As mentioned above, the Milagro collaboration has reported a low-energy cut-off in the spectra of 
the sources MGRO J1908+06, MGRO J2031+41 and MGRO J2019+37~\cite{Abdo:2012jg,Smith:2010yn}. 
In this case, the combinations of $\alpha_\gamma$ and $E_{cut,\gamma}$ that 
could be excluded at 95-99~\%~C.L. using 
future IceCube data are important because they can independently probe the presence of a low-energy cut-off.

\section{\label{sec:conclusion} Conclusions }

The highest energy survey of the Galactic plane has been performed by Milagro. This survey has identified bright sources in the nearby Cygnus star-forming region and in the inner part the Galaxy. Initially, the sources showed the expected behavior of PeVatrons. PeVatrons are the sources of cosmic rays in the "knee" region of the cosmic-ray spectrum that are expected to be sources of pionic gamma rays whose spectrum extends to several hundreds of TeV without a cut-off. Gamma rays from the decay of neutral pions are inevitably accompanied by neutrinos with a flux that is calculable.

In this paper, we re-evaluated the probability of observing four promising Milagro sources in IceCube. We used the updated information from air-Cherenkov and air-shower array experiments to estimate the flux of neutrinos. The prospects for observing these sources in IceCube is highly entangled with discrepancies in the detailed fluxes and morphologies measured by different experiments. Moreover, the uncertainty of the nature of these sources makes it more difficult to understand the observed spectrum. Different spectra and morphology of the sources correspond to different production mechanisms.

It should be noted that the discrepancy between measurements may arise from the difference in angular resolution between air-shower arrays and air-Cherenkov telescopes as well as from the range of energies in which they operate. Future results from HAWC will help resolve these discrepancies and reveal more information about the sources.

If the gamma rays are hadronic in origin, observation of an accompanying neutrino flux is likely over the lifetime of the IceCube experiment. Evidence from IceCube of neutrinos associated with these sources will greatly help in unraveling the nature of the sources.     

\clearpage

\clearpage 

\section*{Acknowledgments}

We would like to thank Markus Ahlers, Chad Finley, M.C.~Gonzalez-Garcia, Zigfried Hampel, Teresa Montaruli, 
Amanda Weinstein, and Tom Weisgarber for useful discussions. 
We would like to specially thank Jay Gallagher, Stefan Westerhoff, and Tova Yoast-Hull for their informative comments and discussions during this study. 
FH and AK research was supported in part by the U.S. National Science Foundation under
Grants No. ANT-0937462 and PHY- 1306958 and by the University of Wisconsin Research 
Committee with funds granted by the Wisconsin Alumni Research Foundation.
VN acknowledges hospitality at WIPAC where part of this work has been carried out. 
VN acknowledges support by Spanish MINECO through project FPA2012-31880, by Spanish MINECO (Centro de excelencia Severo Ochoa Program) under grant SEV-2012-0249. 
VN acknowledges financial support by the European Union through the ITN ELUSIVES H2020-MSCA-ITN-2015//674896 and the RISE INVISIBLESPLUS H2020-MSCA-RISE-2015//690575. 




\section*{References}

\begin{thebibliography}{10}
\expandafter\ifx\csname url\endcsname\relax
  \def\url#1{\texttt{#1}}\fi
\expandafter\ifx\csname urlprefix\endcsname\relax\def\urlprefix{URL }\fi
\expandafter\ifx\csname href\endcsname\relax
  \def\href#1#2{#2} \def\path#1{#1}\fi

\bibitem{halzen:2016}
F.~Halzen, {High-energy neutrino astrophysics}, Nat Phys advance online
  publication.
\newblock \href {http://dx.doi.org/10.1038/nphys3816}
  {\path{doi:10.1038/nphys3816}}.

\bibitem{Taylor:2014hya}
A.~M. Taylor, S.~Gabici, F.~Aharonian, {Galactic halo origin of the neutrinos
  detected by IceCube}, Phys.Rev. D89~(10) (2014) 103003.
\newblock \href {http://arxiv.org/abs/1403.3206} {\path{arXiv:1403.3206}},
  \href {http://dx.doi.org/10.1103/PhysRevD.89.103003}
  {\path{doi:10.1103/PhysRevD.89.103003}}.

\bibitem{Gaggero:2015xza}
D.~Gaggero, D.~Grasso, A.~Marinelli, A.~Urbano, M.~Valli, {The gamma-ray and
  neutrino sky: A consistent picture of Fermi-LAT, Milagro, and IceCube
  results}, Astrophys. J. 815~(2) (2015) L25.
\newblock \href {http://arxiv.org/abs/1504.00227} {\path{arXiv:1504.00227}},
  \href {http://dx.doi.org/10.1088/2041-8205/815/2/L25}
  {\path{doi:10.1088/2041-8205/815/2/L25}}.

\bibitem{Ahlers:2015moa}
M.~Ahlers, Y.~Bai, V.~Barger, R.~Lu, {Galactic neutrinos in the TeV to PeV
  range}, Phys. Rev. D93~(1) (2016) 013009.
\newblock \href {http://arxiv.org/abs/1505.03156} {\path{arXiv:1505.03156}},
  \href {http://dx.doi.org/10.1103/PhysRevD.93.013009}
  {\path{doi:10.1103/PhysRevD.93.013009}}.

\bibitem{Palladino:2016zoe}
A.~Palladino, F.~Vissani, {Extragalactic plus Galactic model for IceCube
  neutrino events}, Astrophys. J. 826~(2) (2016) 185.
\newblock \href {http://arxiv.org/abs/1601.06678} {\path{arXiv:1601.06678}},
  \href {http://dx.doi.org/10.3847/0004-637X/826/2/185}
  {\path{doi:10.3847/0004-637X/826/2/185}}.

\bibitem{Fox:2013oza}
D.~Fox, K.~Kashiyama, P.~M\`eszar\'os, {Sub-PeV Neutrinos from TeV Unidentified
  Sources in the Galaxy}, Astrophys.J. 774 (2013) 74.
\newblock \href {http://arxiv.org/abs/1305.6606} {\path{arXiv:1305.6606}},
  \href {http://dx.doi.org/10.1088/0004-637X/774/1/74}
  {\path{doi:10.1088/0004-637X/774/1/74}}.

\bibitem{Lunardini:2011br}
C.~Lunardini, S.~Razzaque, {High Energy Neutrinos from the Fermi Bubbles},
  Phys.Rev.Lett. 108 (2012) 221102.
\newblock \href {http://arxiv.org/abs/1112.4799} {\path{arXiv:1112.4799}},
  \href {http://dx.doi.org/10.1103/PhysRevLett.108.221102}
  {\path{doi:10.1103/PhysRevLett.108.221102}}.

\bibitem{Lunardini:2013gva}
C.~Lunardini, S.~Razzaque, K.~T. Theodoseau, L.~Yang, {Neutrino Events at
  IceCube and the Fermi Bubbles}, Phys.Rev. D90~(2) (2014) 023016.
\newblock \href {http://arxiv.org/abs/1311.7188} {\path{arXiv:1311.7188}},
  \href {http://dx.doi.org/10.1103/PhysRevD.90.023016}
  {\path{doi:10.1103/PhysRevD.90.023016}}.

\bibitem{Bai:2014kba}
Y.~Bai, A.~J. Barger, V.~Barger, R.~Lu, A.~D. Peterson, J.~Salvado, {Neutrino
  Lighthouse at Sagittarius A*}, Phys. Rev. D90~(6) (2014) 063012.
\newblock \href {http://arxiv.org/abs/1407.2243} {\path{arXiv:1407.2243}},
  \href {http://dx.doi.org/10.1103/PhysRevD.90.063012}
  {\path{doi:10.1103/PhysRevD.90.063012}}.

\bibitem{Adrian-Martinez:2014wzf}
S.~Adrian-Martinez, et~al., {Searches for Point-like and extended neutrino
  sources close to the Galactic Centre using the ANTARES neutrino Telescope},
  Astrophys.J. 786 (2014) L5.
\newblock \href {http://arxiv.org/abs/1402.6182} {\path{arXiv:1402.6182}},
  \href {http://dx.doi.org/10.1088/2041-8205/786/1/L5}
  {\path{doi:10.1088/2041-8205/786/1/L5}}.

\bibitem{Gonzalez-Garcia:2013iha}
M.~Gonzalez-Garcia, F.~Halzen, V.~Niro, {Reevaluation of the Prospect of
  Observing Neutrinos from Galactic Sources in the Light of Recent Results in
  Gamma Ray and Neutrino Astronomy}, Astropart.Phys. 57-58 (2014) 39--48.
\newblock \href {http://arxiv.org/abs/1310.7194} {\path{arXiv:1310.7194}},
  \href {http://dx.doi.org/10.1016/j.astropartphys.2014.04.001}
  {\path{doi:10.1016/j.astropartphys.2014.04.001}}.

\bibitem{Halzen:2008zj}
F.~Halzen, A.~Kappes, A.~O'Murchadha, {Prospects for identifying the sources of
  the Galactic cosmic rays with IceCube}, Phys.Rev. D78 (2008) 063004.
\newblock \href {http://arxiv.org/abs/0803.0314} {\path{arXiv:0803.0314}},
  \href {http://dx.doi.org/10.1103/PhysRevD.78.063004}
  {\path{doi:10.1103/PhysRevD.78.063004}}.

\bibitem{GonzalezGarcia:2009jc}
M.~Gonzalez-Garcia, F.~Halzen, S.~Mohapatra, {Identifying Galactic PeVatrons
  with Neutrinos}, Astropart.Phys. 31 (2009) 437--444.
\newblock \href {http://arxiv.org/abs/0902.1176} {\path{arXiv:0902.1176}},
  \href {http://dx.doi.org/10.1016/j.astropartphys.2009.05.002}
  {\path{doi:10.1016/j.astropartphys.2009.05.002}}.

\bibitem{Aartsen:2014cva}
M.~Aartsen, et~al., {Searches for Extended and Point-like Neutrino Sources with
  Four Years of IceCube Data }\href {http://arxiv.org/abs/1406.6757}
  {\path{arXiv:1406.6757}}, \href
  {http://dx.doi.org/10.1088/0004-637X/796/2/109}
  {\path{doi:10.1088/0004-637X/796/2/109}}.

\bibitem{Abdo:2012jg}
A.~Abdo, U.~Abeysekara, B.~Allen, T.~Aune, D.~Berley, et~al., {Spectrum and
  Morphology of the Two Brightest Milagro Sources in the Cygnus Region: MGRO
  J2019+37 and MGRO J2031+41}, Astrophys.J. 753 (2012) 159.
\newblock \href {http://arxiv.org/abs/1202.0846} {\path{arXiv:1202.0846}},
  \href {http://dx.doi.org/10.1088/0004-637X/753/2/159}
  {\path{doi:10.1088/0004-637X/753/2/159}}.

\bibitem{Smith:2010yn}
A.~J. Smith, {A Survey of Fermi Catalog Sources using Data from the Milagro
  Gamma-Ray Observatory }\href {http://arxiv.org/abs/1001.3695}
  {\path{arXiv:1001.3695}}.

\bibitem{Abeysekara:2015qba}
A.~U. Abeysekara, et~al., {Search for TeV Gamma-Ray Emission from Point-like
  Sources in the Inner Galactic Plane with a Partial Configuration of the HAWC
  Observatory}, Astrophys. J. 817~(1) (2016) 3.
\newblock \href {http://arxiv.org/abs/1509.05401} {\path{arXiv:1509.05401}},
  \href {http://dx.doi.org/10.3847/0004-637X/817/1/3}
  {\path{doi:10.3847/0004-637X/817/1/3}}.

\bibitem{hawc_gamma}
A.~Sandoval, Highlights from hawc, 6th international symposium on high-energy
  gamma-ray astronomy (gamma2016), heidelberg, germany, july 11-15, 2016.

\bibitem{Lemoine:2014ala}
M.~Lemoine, K.~Kotera, J.~P\'etri, {On ultra-high energy cosmic ray
  acceleration at the termination shock of young pulsar winds}, JCAP 1507
  (2015) 016.
\newblock \href {http://arxiv.org/abs/1409.0159} {\path{arXiv:1409.0159}},
  \href {http://dx.doi.org/10.1088/1475-7516/2015/07/016}
  {\path{doi:10.1088/1475-7516/2015/07/016}}.

\bibitem{Abdo:2007ad}
A.~Abdo, B.~T. Allen, D.~Berley, S.~Casanova, C.~Chen, et~al., {TeV Gamma-Ray
  Sources from a Survey of the Galactic Plane with Milagro}, Astrophys.J. 664
  (2007) L91--L94.
\newblock \href {http://arxiv.org/abs/0705.0707} {\path{arXiv:0705.0707}},
  \href {http://dx.doi.org/10.1086/520717} {\path{doi:10.1086/520717}}.

\bibitem{Abdo:2009ku}
A.~Abdo, B.~Allen, T.~Aune, D.~Berley, C.~Chen, et~al., {Milagro Observations
  of TeV Emission from Galactic Sources in the Fermi Bright Source List},
  Astrophys.J. 700 (2009) L127--L131.
\newblock \href {http://arxiv.org/abs/0904.1018} {\path{arXiv:0904.1018}},
  \href {http://dx.doi.org/10.1088/0004-637X/700/2/L127,
  10.1088/0004-637X/703/2/L185} {\path{doi:10.1088/0004-637X/700/2/L127,
  10.1088/0004-637X/703/2/L185}}.

\bibitem{ARGO-YBJ:2012goa}
{Observation of the TeV gamma-ray source MGRO J1908+06 with ARGO-YBJ},
  Astrophys.J. 760 (2012) 110.
\newblock \href {http://arxiv.org/abs/1207.6280} {\path{arXiv:1207.6280}},
  \href {http://dx.doi.org/10.1088/0004-637X/760/2/110}
  {\path{doi:10.1088/0004-637X/760/2/110}}.

\bibitem{Aharonian:2009je}
F.~Aharonian, {Detection of Very High Energy radiation from HESS J1908+063
  confirms the Milagro unidentified source MGRO J1908+06}, Astron.Astrophys.
  499 (2009) 723.
\newblock \href {http://arxiv.org/abs/0904.3409} {\path{arXiv:0904.3409}},
  \href {http://dx.doi.org/10.1051/0004-6361/200811357}
  {\path{doi:10.1051/0004-6361/200811357}}.

\bibitem{Aliu:2014xra}
E.~Aliu, S.~Archambault, T.~Aune, B.~Behera, M.~Beilicke, et~al.,
  {Investigating the TeV Morphology of MGRO J1908+06 with VERITAS},
  Astrophys.J. 787 (2014) 166.
\newblock \href {http://arxiv.org/abs/1404.7185} {\path{arXiv:1404.7185}},
  \href {http://dx.doi.org/10.1088/0004-637X/787/2/166}
  {\path{doi:10.1088/0004-637X/787/2/166}}.

\bibitem{Abdo:2010ht}
A.~Abdo, A.~Abdo, {PSR J1907+0602: A Radio-Faint Gamma-Ray Pulsar Powering a
  Bright TeV Pulsar Wind Nebula}, Astrophys.J. 711 (2010) 64--74.
\newblock \href {http://arxiv.org/abs/1001.0792} {\path{arXiv:1001.0792}},
  \href {http://dx.doi.org/10.1088/0004-637X/711/1/64}
  {\path{doi:10.1088/0004-637X/711/1/64}}.

\bibitem{BaadeAndZwicky}
W.~Baade, F.~Zwicky, On super-novae, Proceedings of the National Academy of
  Sciences 20~(5) (1934) 254--259.
\newblock \href {http://dx.doi.org/10.1073/pnas.20.5.254}
  {\path{doi:10.1073/pnas.20.5.254}}.

\bibitem{Ackermann:2013wqa}
M.~Ackermann, et~al., {Detection of the Characteristic Pion-Decay Signature in
  Supernova Remnants}, Science 339 (2013) 807.
\newblock \href {http://arxiv.org/abs/1302.3307} {\path{arXiv:1302.3307}},
  \href {http://dx.doi.org/10.1126/science.1231160}
  {\path{doi:10.1126/science.1231160}}.

\bibitem{Gabici:2007qb}
S.~Gabici, F.~A. Aharonian, {Searching for galactic cosmic ray pevatrons with
  multi-TeV gamma rays and neutrinos}, Astrophys.J. 665 (2007) L131.
\newblock \href {http://arxiv.org/abs/0705.3011} {\path{arXiv:0705.3011}},
  \href {http://dx.doi.org/10.1086/521047} {\path{doi:10.1086/521047}}.

\bibitem{abdo}
A.~A. Abdo, Discovery of localized tev gamma-ray sources and diffuse tev
  gamma-ray emission from the galactic plane with milagro using a new
  background rejection technique, Ph.D. thesis, Michigan State University,
  Department of Physics and Astronomy (2007).

\bibitem{Bartoli:2012tj}
B.~Bartoli, P.~Bernardini, X.~Bi, C.~Bleve, I.~Bolognino, et~al., {Observation
  of TeV gamma rays from the Cygnus region with the ARGO-YBJ experiment},
  Astrophys.J. 745 (2012) L22.
\newblock \href {http://arxiv.org/abs/1201.1973} {\path{arXiv:1201.1973}},
  \href {http://dx.doi.org/10.1088/2041-8205/745/2/L22}
  {\path{doi:10.1088/2041-8205/745/2/L22}}.

\bibitem{Albert:2008yk}
J.~Albert, et~al., {MAGIC observations of the unidentified TeV gamma-ray source
  TeV J2032+4130}, Astrophys.J. 675 (2008) L25--L28.
\newblock \href {http://arxiv.org/abs/0801.2391} {\path{arXiv:0801.2391}},
  \href {http://dx.doi.org/10.1086/529520} {\path{doi:10.1086/529520}}.

\bibitem{Aharonian:2005ex}
F.~Aharonian, et~al., {The Unidentified TeV source (TeV J2032+4130) and
  surrounding field: Final HEGRA IACT-system results}, Astron.Astrophys. 431
  (2005) 197--202.
\newblock \href {http://arxiv.org/abs/astro-ph/0501667}
  {\path{arXiv:astro-ph/0501667}}.

\bibitem{Lang:2004bk}
M.~J. Lang, D.~Carter-Lewis, D.~Fegan, S.~Fegan, A.~Hillas, et~al., {Evidence
  for TeV gamma ray emission from TeV j2032+4130 in whipple archival data},
  Astron.Astrophys. 423 (2004) 415--419.
\newblock \href {http://arxiv.org/abs/astro-ph/0405513}
  {\path{arXiv:astro-ph/0405513}}, \href
  {http://dx.doi.org/10.1051/0004-6361:20041021}
  {\path{doi:10.1051/0004-6361:20041021}}.

\bibitem{A.WeinsteinfortheVERITAS:2014iwa}
A.~Weinstein, {Pulsar Wind Nebulae and Cosmic Rays: A Bedtime Story},
  Nucl.Phys.Proc.Suppl. 256-257 (2014) 136--148.
\newblock \href {http://arxiv.org/abs/1411.2532} {\path{arXiv:1411.2532}},
  \href {http://dx.doi.org/10.1016/j.nuclphysbps.2014.10.017}
  {\path{doi:10.1016/j.nuclphysbps.2014.10.017}}.

\bibitem{Argo:2014tqa}
B.~Bartoli, et~al., {Identification of the TeV Gamma-ray Source ARGO J2031+4157
  with the Cygnus Cocoon}, Astrophys. J. 790~(2) (2014) 152.
\newblock \href {http://arxiv.org/abs/1406.6436} {\path{arXiv:1406.6436}},
  \href {http://dx.doi.org/10.1088/0004-637X/790/2/152}
  {\path{doi:10.1088/0004-637X/790/2/152}}.

\bibitem{Beacom:2007yu}
J.~F. Beacom, M.~D. Kistler, {Dissecting the Cygnus Region with TeV Gamma Rays
  and Neutrinos}, Phys.Rev. D75 (2007) 083001.
\newblock \href {http://arxiv.org/abs/astro-ph/0701751}
  {\path{arXiv:astro-ph/0701751}}, \href
  {http://dx.doi.org/10.1103/PhysRevD.75.083001}
  {\path{doi:10.1103/PhysRevD.75.083001}}.

\bibitem{Aliu:2014rha}
E.~Aliu, T.~Aune, B.~Behera, M.~Beilicke, W.~Benbow, et~al., {Spatially
  resolving the very high energy emission from MGRO J2019+37 with VERITAS},
  Astrophys.J. 788 (2014) 78.
\newblock \href {http://arxiv.org/abs/1404.1841} {\path{arXiv:1404.1841}},
  \href {http://dx.doi.org/10.1088/0004-637X/788/1/78}
  {\path{doi:10.1088/0004-637X/788/1/78}}.

\bibitem{Kelner:2006tc}
S.~Kelner, F.~A. Aharonian, V.~Bugayov, {Energy spectra of gamma-rays,
  electrons and neutrinos produced at proton-proton interactions in the very
  high energy regime}, Phys.Rev. D74 (2006) 034018.
\newblock \href {http://arxiv.org/abs/astro-ph/0606058}
  {\path{arXiv:astro-ph/0606058}}, \href
  {http://dx.doi.org/10.1103/PhysRevD.74.034018, 10.1103/PhysRevD.79.039901}
  {\path{doi:10.1103/PhysRevD.74.034018, 10.1103/PhysRevD.79.039901}}.

\bibitem{Kappes:2006fg}
A.~Kappes, J.~Hinton, C.~Stegmann, F.~A. Aharonian, {Potential Neutrino Signals
  from Galactic Gamma-Ray Sources}, Astrophys.J. 656 (2007) 870--896.
\newblock \href {http://arxiv.org/abs/astro-ph/0607286}
  {\path{arXiv:astro-ph/0607286}}, \href {http://dx.doi.org/10.1086/508936,
  10.1086/518161} {\path{doi:10.1086/508936, 10.1086/518161}}.

\bibitem{Kappes:2009zza}
A.~Kappes, F.~Halzen, A.~O.~Murchadha, {Prospects of identifying the sources of
  the galactic cosmic rays with IceCube}, Nucl.Instrum.Meth. A602 (2009)
  117--119.
\newblock \href {http://dx.doi.org/10.1016/j.nima.2008.12.049}
  {\path{doi:10.1016/j.nima.2008.12.049}}.

\bibitem{Halzen:2007ah}
F.~Halzen, A.~O~Murchadha, {Neutrinos from Cosmic Ray Accelerators in the
  Cygnus Region of the Galaxy}, Phys.Rev. D76 (2007) 123003.
\newblock \href {http://arxiv.org/abs/0705.1723} {\path{arXiv:0705.1723}},
  \href {http://dx.doi.org/10.1103/PhysRevD.76.123003}
  {\path{doi:10.1103/PhysRevD.76.123003}}.

\bibitem{Vissani:2011ea}
F.~Vissani, F.~Aharonian, {Galactic Sources of High-Energy Neutrinos:
  Highlights}, Nucl.Instrum.Meth. A692 (2012) 5--12.
\newblock \href {http://arxiv.org/abs/1112.3911} {\path{arXiv:1112.3911}},
  \href {http://dx.doi.org/10.1016/j.nima.2011.12.079}
  {\path{doi:10.1016/j.nima.2011.12.079}}.

\bibitem{Vissani:2011vg}
F.~Vissani, F.~Aharonian, N.~Sahakyan, {On the Detectability of High-Energy
  Galactic Neutrino Sources}, Astropart.Phys. 34 (2011) 778--783.
\newblock \href {http://arxiv.org/abs/1101.4842} {\path{arXiv:1101.4842}},
  \href {http://dx.doi.org/10.1016/j.astropartphys.2011.01.011}
  {\path{doi:10.1016/j.astropartphys.2011.01.011}}.

\bibitem{Tchernin:2013wfa}
C.~Tchernin, J.~Aguilar, A.~Neronov, T.~Montaruli, {Neutrino signal from
  extended Galactic sources in IceCube}, Astron.Astrophys. 560 (2013) A67.
\newblock \href {http://arxiv.org/abs/1305.4113} {\path{arXiv:1305.4113}},
  \href {http://dx.doi.org/10.1051/0004-6361/201321801}
  {\path{doi:10.1051/0004-6361/201321801}}.

\bibitem{Honda:2011nf}
M.~Honda, T.~Kajita, K.~Kasahara, S.~Midorikawa, {Improvement of low energy
  atmospheric neutrino flux calculation using the JAM nuclear interaction
  model}, Phys.Rev. D83 (2011) 123001.
\newblock \href {http://arxiv.org/abs/1102.2688} {\path{arXiv:1102.2688}},
  \href {http://dx.doi.org/10.1103/PhysRevD.83.123001}
  {\path{doi:10.1103/PhysRevD.83.123001}}.

\bibitem{Aartsen:2015rwa}
M.~G. Aartsen, et~al., {Evidence for Astrophysical Muon Neutrinos from the
  Northern Sky with IceCube}, Phys. Rev. Lett. 115~(8) (2015) 081102.
\newblock \href {http://arxiv.org/abs/1507.04005} {\path{arXiv:1507.04005}},
  \href {http://dx.doi.org/10.1103/PhysRevLett.115.081102}
  {\path{doi:10.1103/PhysRevLett.115.081102}}.

\bibitem{Alexandreas:1992ek}
D.~Alexandreas, D.~Berley, S.~Biller, G.~Dion, J.~Goodman, et~al., {Point
  source search techniques in ultrahigh-energy gamma-ray astronomy},
  Nucl.Instrum.Meth. A328 (1993) 570--577.
\newblock \href {http://dx.doi.org/10.1016/0168-9002(93)90677-A}
  {\path{doi:10.1016/0168-9002(93)90677-A}}.

\bibitem{ATLAS:2011tau}
{Procedure for the LHC Higgs boson search combination in summer 2011}.

\bibitem{Junk:1999kv}
T.~Junk, {Confidence level computation for combining searches with small
  statistics}, Nucl.Instrum.Meth. A434 (1999) 435--443.
\newblock \href {http://arxiv.org/abs/hep-ex/9902006}
  {\path{arXiv:hep-ex/9902006}}, \href
  {http://dx.doi.org/10.1016/S0168-9002(99)00498-2}
  {\path{doi:10.1016/S0168-9002(99)00498-2}}.

\bibitem{Read:2000ru}
A.~L. Read, {Modified frequentist analysis of search results (The CL(s)
  method)}.

\bibitem{Beringer:1900zz}
J.~Beringer, et~al., {Review of Particle Physics (RPP)}, Phys.Rev. D86 (2012)
  010001.
\newblock \href {http://dx.doi.org/10.1103/PhysRevD.86.010001}
  {\path{doi:10.1103/PhysRevD.86.010001}}.

\end{thebibliography}
\end{document}